\documentclass[prb,twocolumn,showpacs,amsmath,amssymb,superscriptaddress]{revtex4}
\usepackage[dvips]{graphicx}
\usepackage{graphics}
\usepackage{dcolumn}
\usepackage{bm}

\begin{document}
\title{Prospects of high temperature ferromagnetism in (Ga,Mn)As
semiconductors}
\author{T. Jungwirth}
\affiliation{Institute of Physics  ASCR, Cukrovarnick\'a 10, 162 53 Praha 6, Czech Republic }
\affiliation{School of Physics and Astronomy, University of Nottingham, Nottingham NG7 2RD, UK}
\author{K.Y. Wang}
\affiliation{School of Physics and Astronomy, University of Nottingham, Nottingham NG7 2RD, UK}
\author{J. Ma\v{s}ek}
\affiliation{Institute of Physics  ASCR, Na Slovance 2, 182 21 Praha 8, Czech Republic }
\author{K.W. Edmonds}
\affiliation{School of Physics and Astronomy, University of Nottingham, Nottingham NG7 2RD, UK}
\author{J\"{u}rgen K\"{o}nig}
\affiliation{Institut f\"{u}r Theoretische Physik III, Ruhr-Universit\"{a}t Bochum, 44780 Bochum, Germany}
\author{Jairo Sinova}
\affiliation{Department of Physics, Texas A\&M University, College Station,Texas 77843-4242, USA}
\author{M. Polini}
\affiliation{NEST-INFM and Scuola Normale Superiore, I-56126 Pisa, Italy}
\author{N.A. Goncharuk}
\affiliation{Institute of Physics  ASCR, Cukrovarnick\'a 10, 162 53 Praha 6, Czech Republic }
\author{A.H. MacDonald}
\affiliation{Department of Physics, University of Texas at Austin, Austin TX 78712-1081}
\author{M. Sawicki}
\affiliation{Institute of Physics, Polish Academy of Sciences, 02668 Warszawa, Poland}
\author{R.P. Campion}
\author{L.X. Zhao}
\author{C.T. Foxon}
\author{B.L. Gallagher}
\affiliation{School of Physics and Astronomy, University of Nottingham, Nottingham NG7 2RD, UK}
\begin{abstract}
We report on a comprehensive  combined experimental and theoretical study of Curie
temperature trends in (Ga,Mn)As ferromagnetic semiconductors.  Broad agreement
between theoretical expectations and measured data
allows  us to conclude that  $T_c$ in  high-quality metallic samples  
increases  linearly with  the number  of  uncompensated local  moments on  Mn$_{\rm Ga}$
acceptors,  with no  sign  of saturation.  Room temperature 
ferromagnetism is expected for a 10\% concentration of these local moments.   
Our magnetotransport and magnetization data are consistnent with the picture
in which Mn  impurities  
incorporated  during growth at interstitial  Mn$_{\rm I}$ positions act as double-donors and compensate
neighboring Mn$_{\rm Ga}$ local moments because of strong near-neighbor Mn$_{\rm Ga}$-Mn$_{\rm I}$ antiferromagnetic
coupling. These defects can be efficiently removed by post-growth annealing.
Our analysis suggests that there is no fundamental obstacle to
substitutional Mn$_{\rm  Ga}$  doping in high-quality materials beyond our current  maximum level of 6.2\%,
although this achievement will require further advances in growth condition control.
Modest charge compensation does not limit the maximum Curie temperature
possible in ferromagnetic semiconductors based on (Ga,Mn)As.
\end{abstract}

\maketitle

\section{Introduction}

After some frustration in the community 
caused by the difficulties encountered in overcoming the apparent Curie temperature limit in
(Ga,Mn)As of $T_c=110$~K,\cite{Ohno:1998_a,Potashnik:2002_a,Yu:2003_a}
the record  transition temperature  has been steadily  increasing over
the                              last                              two
years.\cite{Edmonds:2002_b,Chiba:2003_b,Ku:2003_a,Edmonds:2004_a,Wang:2004_c}
The maximum  $T_c=173$~K reported\cite{Wang:2004_c} to  date is likely
another short-lived  record in bulk (Ga,Mn)As ferromagnets.  It is now
established  that   the  success  has   been  made  possible   by  the
technological progress  in controlling crystallographic  quality of the
materials, namely, in reducing  the number of unintentional charge and
moment compensating  defects through optimized  growth and post-growth
annealing    procedures.\cite{Chiba:2003_b,Stone:2003_a,Edmonds:2004_a}
Experiments also  suggest that  the general picture  of ferromagnetism
that applies to these metallic (Ga,Mn)As systems is the one in which magnetic coupling
between  local Mn  moments is  mediated  by delocalized  holes in  the
(Ga,Mn)As valence band.  The fact that the mechanism  does not imply a
fundamental $T_c$  limit below  room temperature motivates  a detailed
analysis  of  our  understanding  of  the $T_c$  trends  in  currently
available  high   quality  metallic materials  with  Mn   doping  ranging  from
approximately 2\% to 9\%.

Curie   temperatures   in  metallic   (Ga,Mn)As   have  been   studied
theoretically        starting        from        semi-phenomenological
\cite{Jungwirth:1999_a,Dietl:2000_a,Jungwirth:2002_b,Brey:2003_a,dasSarma:2004_a}
and                                                          microscopic
models\cite{Sandratskii:2002_a,Sato:2003_a,Sandratskii:2004_a,Hilbert:2005_a,Jungwirth:2003_c,Timm:2003_a}
of the  electronic structure. The former approach  asserts a localized
character  of  the five  Mn$_{\rm  Ga}$  d-orbitals  forming a  moment
$S=5/2$  and describes  hole  states  in the  valence  band using  the
Kohn-Luttinger parameterization  for GaAs\cite{Vurgaftman:2001_a} and a
single constant $J_{pd}$  which characterizes the exchange interaction
between Mn$_{\rm  Ga}$ and hole  spins. 
The exchange interaction follows from hybridization between Mn d orbitals and valence band p orbitals.
The semi-phenomenological Hamiltonian implicitly assumes that a canonical transformation has been performed which eliminated the hybridization.\cite{Timm:2003_a} 
 In this approach the hybridization is implicitly assumed to be weak in several different ways, 
 and the canonical transformation ignored in representing observables.  Although this approach is 
 consistent, it should be realized that the localized d-orbitals in the phenomenological Hamiltonian 
 are in reality hybridized with the valence band.  

The advantage
of     the   semi-phenomenological  approach    is     that     it    uses     the experimental
value\cite{Okabayashi:1998_a,Omiya:2000_a}      for      $J_{pd}=54\pm
9$~meV~nm$^3$,  i.e.,  it  correctly  captures  the  strength  of  the
magnetic interaction that has been established to play the key role in
ferromagnetism  in  (Ga,Mn)As.    The  model   also  accounts  for   strong  spin-orbit
interaction present  in the host  valence band which splits  the three
p-bands  into a  heavy-hole,  light-hole, and  a  split-off band  with
different   dispersions.   The   spin-orbit  coupling   is   not  only
responsible        for        a        number       of        distinct
magnetic\cite{Dietl:2001_b,Abolfath:2001_a,Sawicki:2004_b,Sawicki:2004_a}
and
magneto-transport\cite{Jungwirth:2003_b,Tang:2003_a,Gould:2004_a,Giddings:2004_a}
properties of  (Ga,Mn)As ferromagnets but the  resulting complexity of
the valence band was shown\cite{Konig:2001_a,Brey:2003_a} to play also
an  important role  in suppressing  magnetization  fluctuation effects
and, therefore, stabilizing the ferromagnetic state itself.
Describing the potentially complex behavior
of Mn$_{\rm  Ga}$ in GaAs by a  single parameter may oversimplify the problem. The
calculations omit, for example, the contribution of direct antiferromagnetic superexchange
to the coupling of near-neighbor Mn pairs, and  the whole model inevitably breaks down
if valence fluctuations of Mn$_{\rm  Ga}$ d-electrons become strong.

Microscopic theories, whether based on the tight-binding-approximation
(TBA) parameterization of energies  and overlaps of valence orbitals in
the  lattice\cite{Jungwirth:2003_c,Timm:2003_a}  or  on  the  {\em  ab
initio}                local-density-approximation               (LDA)
schemes,\cite{Sandratskii:2002_a,Sato:2003_a,Sandratskii:2004_a,Hilbert:2005_a}
make no  assumption upon  the character of  Mn$_{\rm Ga}$  impurities in
GaAs and their magnetic coupling. They are therefore useful  for studying material trends in $T_c$
as a function  of Mn doping or density and position  in the lattice of
other intentional or unintentional  impurities present  in real systems.\cite{Masek:2004_a}  
Because spin-orbit interactions add to the numerical complexity of calculations 
that are already challenging, they have normally been neglected.  
Another   shortcoming  of   the  {\em ab initio}
approaches is  the incomplete elimination  of self-interaction effects
which leads  to smaller relative  displacement of the Mn  d-levels and
the top of the valence band. This results in an overestimated strength
of the p-d exchange as compared to experiment.

Within the mean-field approximation, which considers thermodynamics of
an isolated  Mn moment in  an effective field and  neglects correlated
Mn-Mn  fluctuations, microscopic calculations\cite{Sandratskii:2004_a}
typically  yield larger   $T_c$'s  than   the  semi-phenomenological
models\cite{Jungwirth:2002_b,Brey:2003_a}  that  use the  experimental
$J_{pd}$  value. Stronger  p-d  exchange in  the microscopic  theories
leads, however, also to a  larger suppression of the Curie temperature
due   to   fluctuation   effects,   especially  so   in   highly-doped
systems.\cite{Sandratskii:2004_a} (A closer agreement in the character
of  the  $T_c$ versus  Mn-doping  curves,  calculated  within the  two
formalisms, is  obtained when  the deficiencies of  density-functional
theories  are partly  eliminated by  introducing a  correlation energy
constant  within the LDA+U  schemes.\cite{Sandratskii:2004_a}) Despite
the   above  weaknesses   of  semi-phenomenological   and  microscopic
calculations, an overall,  qualitatively consistent picture is clearly
emerging  from  these  complementary  theoretical approaches  that,  as  we
discuss  below, provides  a  useful framework  for analyzing  measured
$T_c$'s.

In experimental  Curie temperature studies  it is crucial  to decouple
intrinsic properties of  (Ga,Mn)As ferromagnets from extrinsic effects
due  to the  presence of  unintentional impurities.  Arsenic antisites
(As$_{\rm  Ga}$) and interstitial  manganese (Mn$_{\rm  I}$) represent
two  major  sources  of  charge  compensation in  (Ga,Mn)As  grown  by
low-temperature  molecular  beam  epitaxy  (LT-MBE),  both  acting  as
double-donors\cite{Maca:2002_a,Kudrnovsky:2003_a}.    A  Mn$_{\rm  I}$
cation when  attracted to a  Mn$_{\rm Ga}$ anion compensates  also the
Mn$_{\rm   Ga}$    local   moment   as   the    two   species   are expected to couple
antiferromagnetically\cite{Blinowski:2003_a,Masek:2003_b,Edmonds:2004_a} 
 due  to  superexchange over the whole range from strong to weak charge compensation. 
 
 The
As$_{\rm  Ga}$  antisites are  stable\cite{Bliss:1992_a}  up to  $\sim
450^{\circ}$C which  is well above  the transition temperature  from a
uniform diluted magnetic semiconductor  to a multiphase structure with
metallic  MnAs  and  other  precipitates.  Therefore,  the  number  of
As$_{\rm Ga}$ defects has to be  minimized already during the LT-MBE growth by
precisely     controlling    the     stoichiometry     of    deposited
epilayers.\cite{Campion:2003_b}   The  Mn$_{\rm  I}$   impurity concentration
can be significant in as-grown structures. These defects are,  however, much more mobile
than           the            As           antisites.           During
annealing\cite{Yu:2002_a,Stone:2003_a,Edmonds:2004_a}  at temperatures
close  to   the  MBE  growth  temperature   $\sim  200^{\circ}$C  they
out-diffuse and are passivated at  the epilayer surface. In this paper
we have collected data for a set of samples that show very weak charge
and moment  compensation after annealing, i.e. a  negligible number of
As$_{\rm  Ga}$,  which allows  us  to  determine experimentally  $T_c$
trends related to intrinsic properties of (Ga,Mn)As ferromagnets.

The paper is organized  as follows: In the theory Section~\ref{theory}
we  start  with  the  semi-phenomenological  mean-field  approximation
(Section~\ref{theory_mf})  to  set  up   a  scale  of  expected  Curie
temperatures  in the material,  assuming homogeneous distribution of Mn$_{\rm Ga}$
ions  (the  virtual-crystal approximation).  We  then discuss  various
physically   distinct   effects   that are not captured by this   picture.   In
Section~\ref{theory_stoner}  we evaluate  the Stoner  enhancement of  the
Curie temperature  due to hole-hole exchange  interaction. Suppression of
$T_c$ due to  
antiferromagnetic superexchange contribution to the near-neighbor Mn$_{\rm Ga}$-Mn$_{\rm Ga}$ coupling in highly compensated samples\cite{Kudrnovsky:2004_a}
is  illustrated  in  Section~\ref{theory_tba}.
In this section we discuss also effects on $T_c$ arising from the discreteness of random Mn$_{\rm Ga}$
positions in the lattice that becomes important in the opposite regime, i.e., in systems with low
charge compensation or co-doped with additional non-magnetic acceptors.
Effects beyond the  mean-field approximation, namely the disappearance
of  the ferromagnetic long-range  order due  to collective  Mn$_{\rm  Ga}$ moments
fluctuations  are  discussed  in  Section~\ref{theory_sw}.  Since  the
Mn$_{\rm  Ga}$,  Mn$_{\rm  I}$,   and  hole  densities  represent  key
parameters in the discussion of measured Curie temperatures, we present
in    Section~\ref{theory_partial}    theoretical   predictions    for
equilibrium partial concentrations of substitutional Mn$_{\rm Ga}$ and
interstitial  Mn$_{\rm  I}$ impurities  in  as-grown  samples, and  in
Section~\ref{theory_hall}  we  estimate   the  accuracy  of  the  Hall
measurement of hole density in the polarized (Ga,Mn)As valence bands.

Measured $T_c$  and hole densities are  presented in Section~\ref{exp_tc}
for  a set  of samples  with different  nominal Mn-doping,  before and
after  annealing.   Motivated by  the  above  theoretical analysis  we
determine  in Section~\ref{exp_partial} the  partial density  of Mn$_{\rm
Ga}$  and Mn$_{\rm  I}$, and  the effective  density  of uncompensated
Mn$_{\rm  Ga}$ local  moments in  our samples.  The  interpretation is
based  on total  Mn-doping values,  obtained from  secondary  ion mass
spectroscopy  (SIMS),  and Hall  measurements  of  the hole  densities
before and after annealing. Consistency of these results is checked by
comparisons   with    independent   magnetization   measurements.   In
Section~\ref{exp_theory} we present experimental $T_c$ dependencies on
uncompensated Mn$_{\rm Ga}$ moment  and hole densities and compare the
data  with theory  predictions.  Technological issues  related to  the
growth  of (Ga,Mn)As epilayers  with large  Mn concentrations  are 
discussed in Section~\ref{discussion}.
Our
perspective   on    high-temperature   ferromagnetism   in   (Ga,Mn)As
semiconductors is summarized in Section~\ref{conclusion}. 

\section{Theory}
\label{theory}

\subsection{Mean-field virtual crystal approximation}
\label{theory_mf}

The description  of ordered states in (Ga,Mn)As  is greatly simplified
by the virtual-crystal approximation in which the random
Mn$_{\rm Ga}$  distribution is replaced  by a continuum with  the same
average  local  moment  density  and  the role  of  other  defects  is
neglected,  apart  from the  potential  hole or  moment compensation.
\cite{Dietl:2000_a,Dietl:1997_a,Jungwirth:1999_a,Konig:2003_a}
Microscopic TBA calculations showed\cite{Jungwirth:2003_c} very little
effect of positional disorder on the strength of magnetic couplings in
(Ga,Mn)As  epilayers with  metallic conductivities  of  interest here,
which  partly justifies the  virtual-crystal approach.  Other detailed
theoretical studies, corroborated  by experimental data below, confirm
the  absence   of  any   significant  magnetic  frustration   in  this
ferromagnetic  semiconductor associated with  the random  positions of
Mn$_{\rm Ga}$ moments in the lattice.\cite{Timm:2004_b,Fiete:2004_a}

In  the  mean-field approximation\cite{Jungwirth:1999_a,Dietl:2000_a},
each local Mn$_{\rm  Ga}$  moment is described by a  Hamiltonian $\vec S_{I}
\cdot \vec H_{MF}$  where $\vec S_{I}$ is the  Mn$_{\rm  Ga}$ local spin operator,
$\vec H_{MF}  = J_{pd}  \langle {\vec s}\rangle$,  and $\langle
{\vec  s}\rangle$  is  the  mean  spin density  of  the  valence  band
holes. $H_{MF}$ is an effective field seen by the local moments due to
spin-polarization of  the band holes, analogous to  the nuclear Knight
shift.   Similarly   $\vec   h_{MF}   =  J_{pd}   N_{Mn}   \langle\vec
S\rangle $  is an effective  magnetic field experienced  by the
valence band  holes, where  $\langle\vec S\rangle $  is the  mean spin
polarization of the Mn$_{\rm  Ga}$ local moments, and $N_{\rm Mn}=4x/a_{\rm lc}^3$
is  the Mn$_{\rm  Ga}$ density  in Ga$_{1-x}$Mn$_x$As  with  a lattice
constant $a_{\rm  lc}$. The dependence  of $\langle\vec S\rangle  $ on
temperature  and  field $H_{MF}$  is  given\cite{Konig:2003_a} by  the
Brillouin function:
\begin{equation}
\langle\vec        S\rangle=       \frac{\vec       H_{MF}}{|H_{MF}|}S
B_s(S|H_{MF}|/k_BT)\; . 
\label{bs}
\end{equation}
The  Curie temperature  is  found by  linearizing  $H_{MF}$ and  $B_s$
around $\langle\vec S\rangle =0$:
\begin{eqnarray}
\vec H_{MF}&\approx& J_{pd}^2N_{Mn} \langle\vec S\rangle\chi_f
\nonumber \\
B_s&\approx&\frac{S+1}{3}\frac{S|H_{MF}|}{k_BT_c}\; . 
\label{linear}
\end{eqnarray}
Here $\chi_f$ is the itinerant hole spin susceptibility given by
\begin{equation}
\chi_f=\frac{d\langle s\rangle}{dh_{MF}} =-\frac{d^2e_T}{dh_{MF}^2}\; , \label{chi}
\end{equation}
and $e_T$ is the total energy per volume of the holes. Eqs.~(\ref{bs}) 
and (\ref{linear}) give
\begin{equation}
  k_B T_c = \frac{N_{\rm Mn} S (S+1)}{3}
  J_{\rm pd}^2\chi_f \; .
\label{tc}
\end{equation}

Qualitative    implications     of    this
$T_c$-equation~(\ref{tc}) can be understood within  a  model
itinerant hole system  with a single spin-split band  and an effective
mass $m^{\ast}$. The kinetic  energy contribution, $e_k$, to the total
energy of the band holes gives a susceptibility:
\begin{equation}
\chi_{f,k}=\frac{d^2e_k}{dh_{MF}^2}= \frac{m^{\ast}k_F}{4\pi^2\hbar^2}
  \; ,
\label{chik}
\end{equation}
where $k_F$  is the Fermi wavevector. Within  this approximation $T_c$
 is  proportional to  the Mn$_{\rm  Ga}$  density, to  the hole  Fermi
 wavevector, i.e. to  $p^{1/3}$ where $p$ is the  hole density, and to
 the hole effective mass $m^{\ast}$.

To obtain quantitative predictions for the critical temperature, it is
necessary  to  evaluate  the  itinerant hole  susceptibility  using  a
realistic band Hamiltonian, $H=H_{KL}+ \vec s\cdot \vec h_{MF}$,
where  $H_{KL}$ the  six-band Kohn-Luttinger  model of  the  GaAs host
band\cite{Vurgaftman:2001_a}   and   $\vec  s$   is   the  hole   spin
operator.\cite{Dietl:2000_a,Dietl:2001_b,Abolfath:2001_a} The results,
represented  by  the  solid  black line  in  Fig.~\ref{tc_mf_sw},  are
consistent with  the qualitative analysis based on  the parabolic band
model, i.e,  $T_c$ follows roughly the $\sim xp^{1/3}$  dependence. Based on
these  calculations, room temperature  ferromagnetism in  (Ga,Mn)As is
expected for 10\% Mn$_{\rm Ga}$ doping in weakly compensated samples.

\subsection{Stoner enhancement of ${\bf T_c}$}
\label{theory_stoner} 
In highly doped (Ga,Mn)As epilayers the hole-hole correlation effects are weak and can
be neglected. The  exchange total energy $e_x$ adds  a contribution to
the hole spin susceptibility:
\begin{equation}
\chi_{f,x}=-\frac{d^2e_x}{dh_{MF}^2}
\; , \label{chitx}
\end{equation}
which for a single parabolic spin-split band reads,
\begin{equation}
 \chi_{f,x}=
  \frac{e^2 (m^{\ast})^2}{4\pi^3\varepsilon \hbar^4} \; ,
\label{chix}
\end{equation}
where   $\varepsilon$  is   the  dielectric   constant  of   the  host
semiconductor.  Eq.~(\ref{chix}) suggests  a  hole-density independent
Stoner   enhancement   of   $T_c$   proportional  to   Mn$_{\rm   Ga}$
concentration and $(m^{\ast})^2$.

\begin{figure}
\includegraphics[width=3.8in]{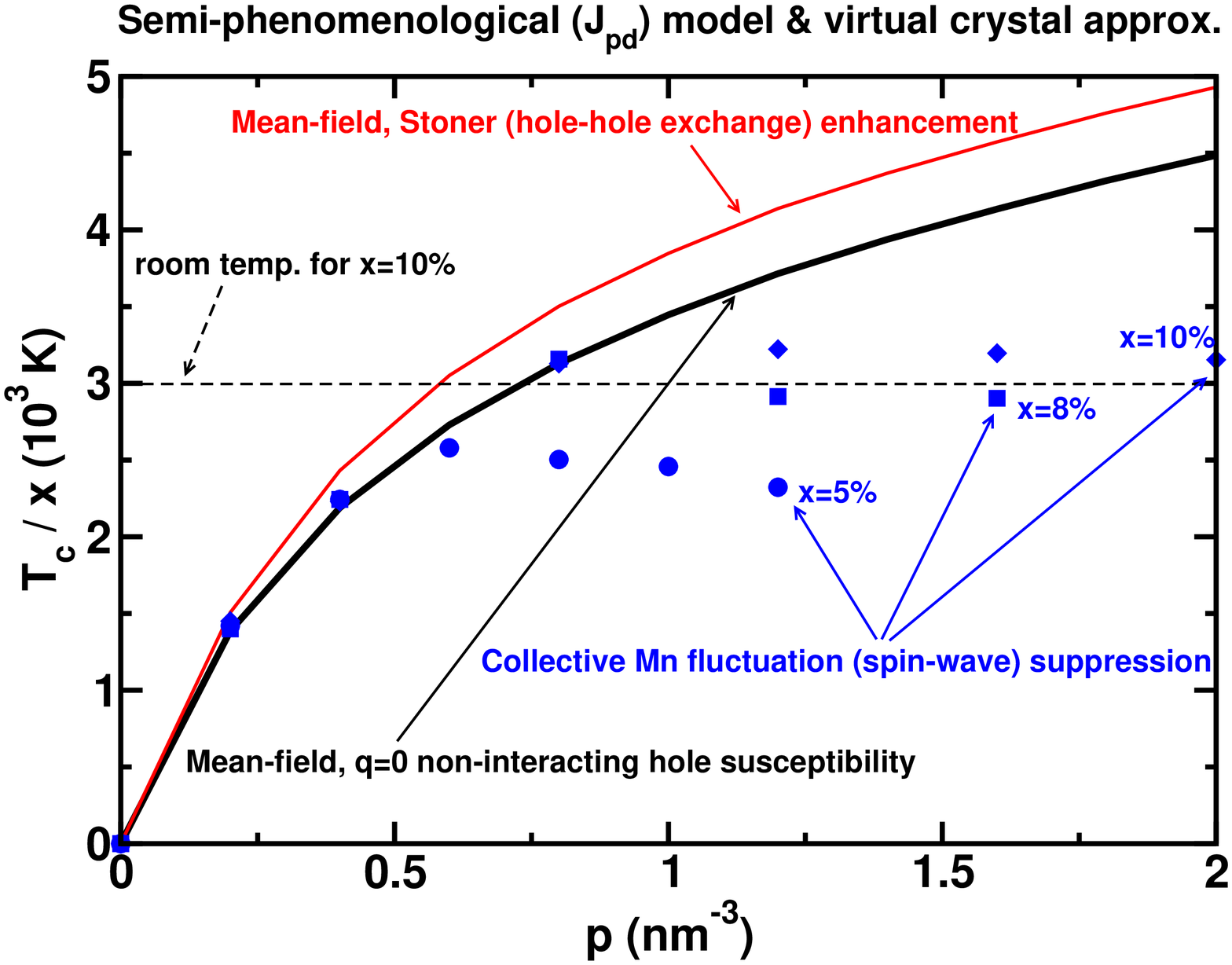}
\caption{Ferromagnetic transition temperatures of (Ga,Mn)As calculated within the
effective Hamiltonian and virtual crystal approximation: 
mean-field (thick black line), Stoner enhancement of $T_c$ (thin red line), spin-wave
suppression of $T_c$ (blue symbols). 
} 
\label{tc_mf_sw}
\end{figure}

As in the non-interacting hole case discussed above, the detailed
valence-band structure  has to be  accounted for to  make quantitative
estimates  of   the  Stoner  $T_c$   enhancement.  The  red   line  in
Fig.~\ref{tc_mf_sw}  shows  the  Stoner $T_c$  enhancement  calculated
numerically from Eqs.~(\ref{chitx}).  As expected, $T_c$ stays roughly
proportional to $xp^{1/3}$ even  if hole-hole exchange interactions are
included,  and  the  enhancement  of  the  Curie  temperature  due  to
interactions is of order $\sim 10-20$\%.

\subsection{Discreteness of random Mn$_{\rm\bf Ga}$ positions, superexchange}
\label{theory_tba} 
So far, the mean-field analysis of $T_c$ has neglected  discreteness in
the random Mn$_{\rm  Ga}$ positions in the lattice and other magnetic coupling mechanisms, 
beside the kinetic-exchange, in particular the near neighbor superexchange. 
The former point can be expected to affect $T_c$ at large hole densities, i.e., when the hole Fermi
wavelength approaches inter-atomic distances.
In the opposite limit of strongly compensated systems, where the overall
magnitude
of the hole-mediated exchange is weaker, 
antiferromagnetic superexchange
can dominate  the near-neighbor Mn$_{\rm  Ga}$-Mn$_{\rm  Ga}$ coupling,\cite{Kudrnovsky:2004_a}  
leading  to a reduced
Curie temperature.\cite{Jungwirth:2003_c,Sandratskii:2004_a} 
This type of magnetic
interaction was ignored in the previous section. 
We emphasize that the phenomenological model cannot be applied consistently 
when nearest neighbor interactions dominate, 
since it implicitly assumes that all length scales are longer than a lattice constant.  
We also note that net antiferromagnetic coupling of near-neighbor Mn$_{\rm Ga}$-Mn$_{\rm Ga}$ 
pairs is expected only in systems with large charge compensations. In weakly compensated (Ga,Mn)As the ferromagnetic contribution takes over.\cite{Kudrnovsky:2004_a,Mahadevan:2004_c}

Beside  the above effects of random Mn distribution, Mn positional disorder can directly modify  p-d 
interaction when the coherence of  Bloch states becomes significantly
disturbed. Microscopic theories, such as the TBA/CPA 
calculations\cite{Jungwirth:2003_c} 
presented in this section or approaches based on
{\em ab-initio} LDA band structure \cite{Sandratskii:2004_a}, 
capture all these effects on an equal footing and can be used 
to estimate trends in mean-field $T_c$ beyond the virtual crystal
approximation.   
The
theories  do  not assert  any  specific  magnetic
coupling  mechanism from the  outset. Instead,  these follow  from the
microscopic  description  of the  electronic  structure  of the  doped
crystal.   

Within the CPA,
disorder effects  appear in  the finite  spectral width  of  hole
quasiparticle states.  
Since realizations with  near-neighbor Mn$_{\rm Ga}$ ions are
included within the disorder-averaged  TBA/CPA with the proper statistical
probability, short-range local moment  interactions (such as superexchange)
contribute to the final
magnetic  state.

The parameterization  of  our TBA Hamiltonian was chosen to provide  the correct band gap
for  a  pure  GaAs  crystal \cite{Talwar:1982_a}  and  the  appropriate
exchange splitting of  the Mn d--states. Local changes  of the crystal
potential at Mn$_{\rm Ga}$, represented
by     shifted     atomic      levels,     are     estimated     using
Ref.~\onlinecite{Harrison:1980_a}.   Long-range   tails   of  the   impurity
potentials,  which  become less  important  with  increasing level  of
doping, are  neglected. (Note, that the  Thomas-Fermi screening length
is     only     3-5${\rm \AA}$     for    typical     carrier     densities
\cite{Jungwirth:2002_c},  i.e., comparable  to the  lattice constant.)
Also lattice relaxation effects are neglected within the CPA.

In our TBA/CPA calculations, hole density is varied independently of 
Mn$_{\rm Ga}$ doping
by adding non-magnetic donors (Si or Se) or acceptors (C or Be).
The resulting valence-band splitting is almost independent of the
density of these non-magnetic impurities at fixed $N_{Mn}$, 
which indicates that  quasiparticle
broadening due to positional disorder has only a weak effect on the
strength of the kinetic-exchange coupling. 
We intentionally  did not use Mn$_{\rm I}$ donors in these calculations
to avoid mixing
of the (arguably) most important effect of this defect which is  moment
compensation. This is discussed separately below in 
Section~\ref{theory_partial}.

The TBA/CPA Curie temperatures are obtained  using  the
compatibility  of  the model with  the  Weiss  mean-field theory.   The
strength of the Mn$_{\rm Ga}$-Mn$_{\rm Ga}$ coupling  is characterized by the energy cost of
flipping one Mn$_{\rm Ga}$ moment, which can be calculated for a given
chemical composition.\cite{Masek:1991_a} 
This effective field $H_{eff}$ 
corresponds to $H_{MF}$  in the semi-phenomenological kinetic-exchange
model used in the previous section, i.e.,

\begin{equation}
k_BT_c=\frac{S+1}{3}H_{eff} \; .
\label{tc_heisenberg}
\end{equation}

\begin{figure}[h]
\includegraphics[width=3.8in]{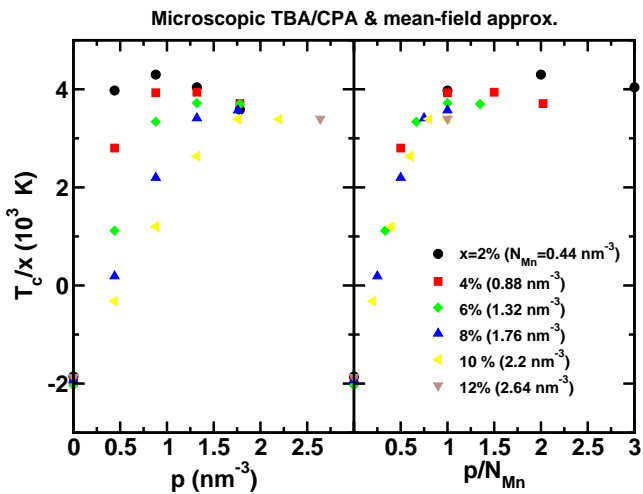}
\caption{$T_c$ calculations within the microscopic TBA/CPA model: $T_c$ versus hole density 
(left panel), $T_c$ versus number of holes per  Mn$_{\rm Ga}$ (right panel).} 
\label{tc_tba}
\end{figure}

In Fig.~\ref{tc_tba} we plot the mean-field TBA/CPA transition
temperatures as a function
of hole densities for several Mn$_{\rm Ga}$ concentrations. 
Since the typical $T_c$'s here are similar to those in Fig.~\ref{tc_mf_sw}
we can identify, based on the comparison between the two figures, the main
physical origins of the deviations from the $T_c\sim xp^{1/3}$ trend.
Black dots in the left panel of Fig.~\ref{tc_tba}
which correspond to a relatively low local Mn$_{\rm Ga}$ moment 
concentration ($x=2\%$) and hole densities ranging up to $p=4N_{Mn}$ show
the expected supression of $T_c$ at large $p$. The effect of superexchange
in the opposite limit is clearly seen when inspecting, e.g., the $x=10\%$ data
for $p<1$~nm$^{-3}$. The mean-field TBA/CPA 
Curie temperature is largely suppressed
here or even negative, meaning that the ferromagnetic state becomes unstable
due to the short-range antiferromagnetic coupling. 
Also, the 
inhomogeneity of the carrier distribution in the disordered mixed 
crystal may contribute to the steep decrease of $T_c$ with increasing 
compensation.
Although
the Curie temperatures in the left panel of Fig.~\ref{tc_tba} appear to
depart strongly for the $T_c\sim xp^{1/3}$ dependence, the linearity
with $x$ is almost fully recovered when
$T_c$ is plotted as a function of the number of holes per Mn$_{\rm Ga}$,  $p/N_{Mn}$
(see right panel of Fig.~\ref{tc_tba}). Note that for compensations ($1-p/N_{Mn}$) reaching 100\%
this property of the superexchange coupling is reminiscent of the behavior
of (II,Mn)VI diluted magnetic semiconductors \cite{Furdyna:1988_b} in which Mn acts as an isovalent
magnetic impurity. The dependence on $p$ in (Ga,Mn)As is expected
to become very weak, however, when reaching the uncompensated limit or when further
increasing hole density by non-magnetic acceptor co-doping. 

\subsection{Collective Mn-moments fluctuations}
\label{theory_sw} 
The potential importance of correlated Mn-moment fluctuations on $T_c$ in (Ga,Mn)As can be illustrated
by recalling, within a simple parabolic band model, the RKKY (or Friedel) oscillations effect which
occurs as a consequence
of the $2k_F$ anomaly in the wavevector dependent susceptibility of the 
hole system.\cite{Dietl:1997_a,Brey:2003_a} 
In this picture, the sign of the indirect kinetic-exchange 
Mn$_{\rm  Ga}$-Mn$_{\rm  Ga}$ coupling fluctuates
as $\cos(2k_Fd)$, where $d$ is the distance between Mn$_{\rm  Ga}$ moments, and its
amplitude decays as $d^3$. 
We can estimate the average Mn$_{\rm  Ga}$-Mn$_{\rm  Ga}$ separation in
a (Ga,Mn)As random alloy as
$\bar{d}=2(3/4\pi N_{Mn})^{1/3}$. If the spin-orbit interaction
and band-warping are neglected, the top of the valence band is formed by
six degenerate parabolic bands with $k_F=(\pi^2p)^{1/3}$. For uncompensated
(Ga,Mn)As systems ($p=N_{Mn}$), we then obtain 
$\cos(2k_F\bar{d})\approx -1$ which means that role of these fluctuations cannot be generally discarded. 
In realistic valence bands, as we see below, the  fluctuations is suppressed due to  non-parabolic and
anisotropic dispersions of the heavy- and light-hole bands and due to the strong spin-orbit coupling.

On a more quantitative level, we   can   establish  the   range   of   reliability   and estimate corrections
to  the   
mean-field theory in  (Ga,Mn)As by accounting for the
suppression
of   the   Curie   temperature  within
quantum theory of long-wavelength   spin-waves
in the semi-phenomenological virtual-crystal model.
We note that a
qualitatively similar picture is obtained using Monte Carlo simulations
which treat Mn-moments as classical variables and account for positional
disorder. \cite{Schliemann:2001_b,Konig:2003_a,Brey:2003_a}
Isotropic ferromagnets have spin-wave Goldstone collective modes whose
energies vanish at long wavelengths,
\begin{equation}
\label{long-wavelength}
   \Omega_{k} = D k^2 + {\cal O} (k^4)\, ,
\end{equation}
where $k$  is the wavevector  of the mode. Spin-orbit  coupling breaks
rotational symmetry and leads to  a finite gap. According to numerical
studies,\cite{Konig:2001_a}  this gap is  small however,  much smaller
than $k_BT_c$ for  example, and plays a role  in magnetic fluctuations
only at very low temperatures.  Spin-wave excitations reduce the total
spin by one, at an energy  cost that is, at least at long wavelengths,
much  smaller   than  the   mean-field  value, $H_{MF}$.  
The  importance  of  these  correlated  spin  excitations,
neglected  by  mean-field  theory,  can  be judged  by  evaluating  an
approximate $T_c$ bound  based on the following argument  which uses a
Debye-like model for the  magnetic excitation spectrum. When spin-wave
interactions  are   neglected,  the  magnetization   vanishes  at  the
temperature where  the number of  excited spin waves equals  the total
spin of the ground state:
\begin{equation}
   N_{\rm Mn}S = \frac{1}{2 \pi^2} \int_0^{k_D} dk\, k^2 n(\Omega_{k}) \, ,
\end{equation}
where $n(\Omega_{k})$ is the Bose occupation number and the Debye cutoff, 
$k_D=(6\pi^2N_{\rm Mn})^{1/3}$. It
follows that the ferromagnetic transition temperature cannot exceed
\begin{equation}
   k_BT_c = \frac{2S+1}{6}  k_D^2 D(T_c) \; .
\label{collective_tc}
\end{equation}
In applying this formula to estimate $T_c$ we have approximated the 
temperature dependence of the spin stiffness
by
\begin{equation}
D(T)=D_0\langle S\rangle(T)/S\; , \label{dt}
\end{equation}
where $D_0$ is  the zero-temperature stiffness,\cite{Konig:2001_a,Schliemann:2001_a} and
$\langle       S\rangle(T)$      is       the       mean-field      Mn
polarization\cite{Abolfath:2001_a}  at  a   temperature  $T$.  If  the
difference between $T_c$ obtained from the self-consistent solution of
Eqs.~(\ref{collective_tc})  and (\ref{dt})  and  the mean-field  Curie
temperature in Eq.~(\ref{tc}) is large, the typical local valence-band
carrier polarization will remain finite above the critical temperature
and  ferromagnetism  will  disappear  only  because  of  the  loss  of
long-range spatial order, the  usual circumstance for transition metal
ferromagnetism for example.

In  discussing corrections  to mean-field-theory  $T_c$  estimates, we
compare spin-stiffness  results obtained with the  simple two-band and
realistic six-band models. Details  on the formalism used to calculate
$D_0$ can be found in Refs.~\onlinecite{Konig:2000_a,Konig:2001_a}. We
find that the zero-temperature spin stiffness is always much larger in
the six-band  model. For (Ga,Mn)As, the  two-band model underestimates
$D_0$  by a factor  of $\sim$10-30  over the  range of  hole densities
considered.   Furthermore, the  trend  is different:  in the  two-band
model the  stiffness decreases with increasing density,  while for the
six-band   description  the   initial  increase   is  followed   by  a
saturation. Even in the limit of low carrier concentrations, it is not
only the (heavy-hole)  mass of the lowest band  which is important for
the  spin stiffness.  In  the realistic  band model,  heavy-holes have
their spin and orbital angular momenta aligned approximately along the
direction  of the  Bloch  wavevector.  Exchange  interactions with  Mn
spins  mix the  heavy  holes  with more  dispersive  light holes.  The
calculations  show  that heavy-light  mixing  is  responsible for  the
relatively large spin stiffnesses.  Crudely, the large mass heavy-hole
band  dominates the  spin  susceptibility and  enables local  magnetic
order  at  high temperatures,  while  the  dispersive light-hole  band
dominates   the  spin  stiffness   and  enables   long-range  magnetic
order.  The multi-band  character  of the  semiconductor valence  band
plays an important role in the ferromagnetism of these materials.

Blue  symbols in  Fig.~\ref{tc_mf_sw}  summarize critical  temperature
estimates that  include both the  Stoner enhancement of $T_c$  and the
suppression due  to spin-wave  fluctuations. The data  were calculated
using the six-band Kohn-Luttinger model  for hole densities up to one
hole per  Mn$_{\rm Ga}$ and Mn$_{\rm Ga}$  concentrations $x=5,8$~ and
10\%. Given  the qualitative  nature of these  $T_c$ estimates  we can
conclude that  $T_c$ will remain  roughly proportional to $x$  even at
large dopings.  The suppression of  $T_c$ due to  spin-waves increases
with   increasing  hole   density   relative  to   the  local   moment
concentration, resulting in saturation of the critical temperature
with increasing $p$ at about 50\% compensation.   

\subsection{${\rm\bf Mn_{Ga}}$ and ${\rm\bf Mn_I}$ partial concentrations}
\label{theory_partial}

In the previous sections we have considered Mn to occupy only the Ga
substitutional positions and found that $T_c$ should increase
linearly with the concentration of Mn$_{\rm Ga}$ local moments. 
In real (Ga,Mn)As materials a fraction of
Mn is incorporated during the growth in interstitial positions. These 
donor impurities are likely to form pairs with Mn$_{\rm Ga}$ acceptors
in as-grown systems with approximately zero net moment,\cite{Blinowski:2003_a,Masek:2003_b,Edmonds:2004_a} 
resulting in an effective free local-moment doping $x_{eff}=x_{s}-x_{i}$.
Here   $x_{s}$  and $x_{i}$ are partial  concentrations   of
substitutional and interstitial Mn, respectively. Although
Mn$_{\rm I}$ can be removed by low-temperature annealing, $x_{eff}$ will
remain smaller than the total nominal Mn doping. The Mn$_{\rm Ga}$ doping
efficiency is, therefore, one of the key parameters that may limit maximum
$T_c$ achieved in (Ga,Mn)As epilayers.

In this Section, we  calculate cohesion energy $E_{c}(x_{s},x_{i})$ as
a  function  of the  partial  concentrations  $x_{s}$  and $x_{i}$   
and use
it to determine the dependence of  $x_{s}$ and $x_{i}$ on the total Mn
doping in as-grown  materials. We define $E_{c}(x_{s},x_{i})$ as a
difference of the crystal energy per unit cell and a properly weighted
sum of energies of isolated  constituent atoms. The cohesion energy is
not very sensitive to the  details of the electronic structure and can
be calculated  with a  reasonable accuracy, for  example by  using the
microscopic TBA model. Note that the growth kinetics calculations\cite{Erwin:2002_a}  
identified adsorption pathways for  Mn$_{\rm I}$ formation in (Ga,Mn)As epilayers. Our equilibrium 
consideration provide, as seen in Section~\ref{exp_partial}, a very good estimate for the 
fraction of Mn impurities incorporated in interstitial positions. 

The  partial Mn concentrations  $x_{s}$ and  $x_{i}$ can  be  obtained by
minimizing $E_{c}(x_{s},x_{i})$ at fixed Mn concentration $x = x_{s} +
x_{i}$  with  respect to  either  $x_{s}$  or  $x_{i}$. Formally,  the
condition for a dynamical equilibrium  between the two positions of Mn
has a form
\begin{equation}
\frac{\partial E_{c}(x_{s},x_{i})}{\partial x_{s}} - 
\frac{\partial E_{c}(x_{s},x_{i})}{\partial x_{i}} = 0.
\label{cohesion_cond}
\end{equation}
It was recently shown \cite{Masek:2002_a} that the partial derivatives
of the  cohesion energy  $E_{c}(x_{s},x_{i})$ with respect  to $x_{s}$
and $x_{i}$ represent formation energies $F_{s}$ and $F_{i}$
of Mn$_{\rm  Ga}$ and Mn$_{\rm I}$  impurities, respectively, assuming
that the atomic  reservoir is formed by neutral  isolated atoms. 
The equilibrium distribution  of
Mn$_{\rm Ga}$ and Mn$_{\rm I}$  is therefore reached 
when 
\begin{equation}
F_{s}(x_{s},x_{i}) = F_{i}(x_{s},x_{i}),
\end{equation}
as  expected also  from  the growth  point  of  view.  Partial
concentrations $x_{s,i}$  of Mn can  be obtained by solving  
Eq.~(\ref{cohesion_cond})
together   with   the  condition   $0   \leq   x_{s,i}   \leq   x$,
$x_{s}+x_{i}=x$.

\begin{figure}
\includegraphics[angle=0,width=3.8in]{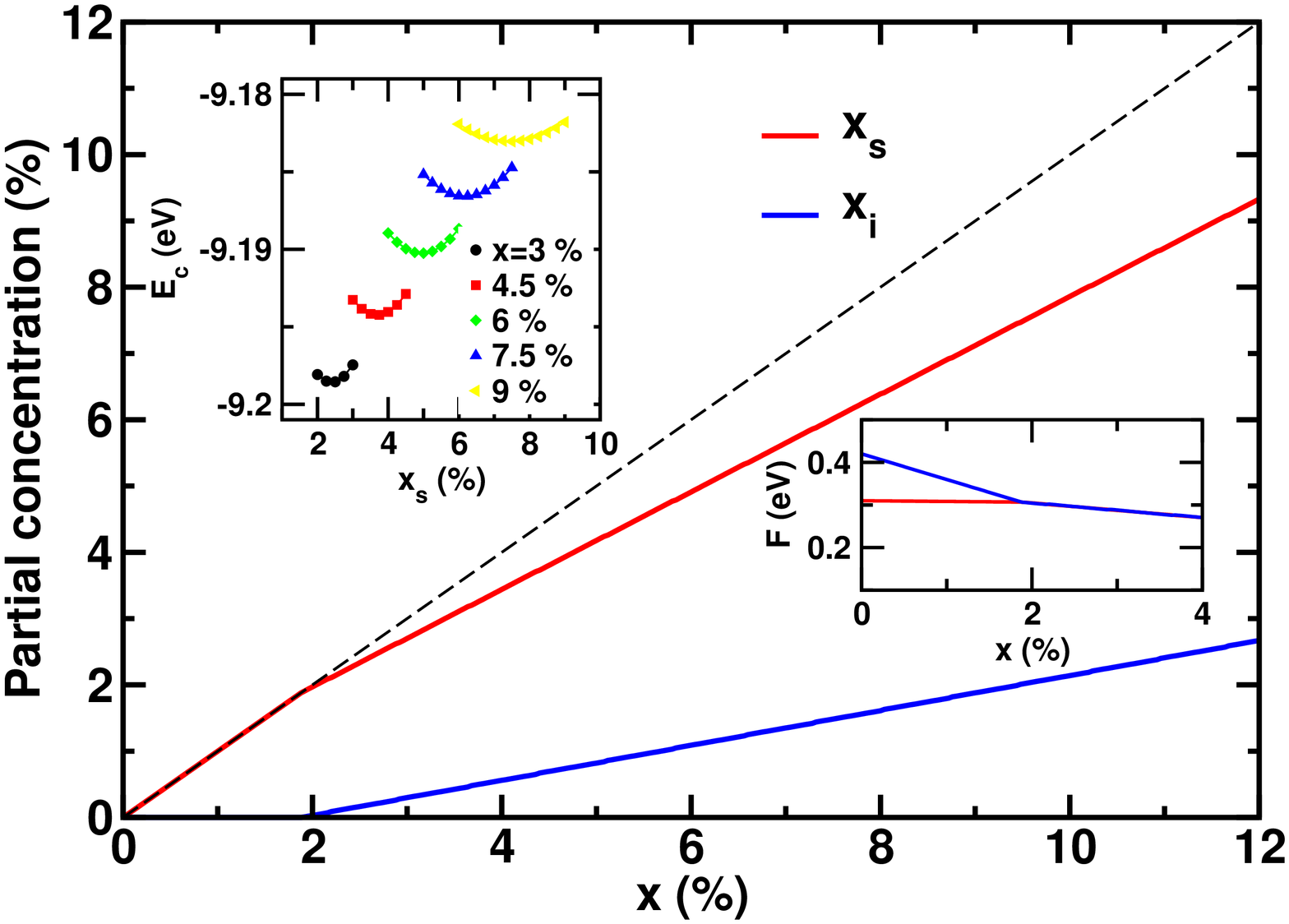}
\caption{
Main panel: Theoretical equilibrium partial concentrations of substitutional Mn$_{\rm Ga}$ (red line) and interstitial
Mn$_{\rm I}$ (blue line) impurities. Right inset: Formation energies of Mn$_{\rm Ga}$ and Mn$_{\rm I}$ as a function
of total Mn concentration.
Left inset: Cohesion energy as a function of substitutional  
Mn$_{\rm Ga}$ concentration at several fixed total Mn concentrations. 
}
\label{partial}
\end{figure}

The left inset of Fig.~\ref{partial}  summarizes the  compositional  dependence of  the
cohesion energy in (Ga,Mn)As with  both Mn$_{\rm Ga}$ and Mn$_{\rm I}$
impurities. We consider several values of  $x$ and
plot $E_{c}(x_{s},x-x_{s})$ vs $x_{s}$. 
Although  the changes of the cohesion energy
due to the incorporation of Mn are small, a systematic linear shift of
the     minimum    of     $E_{c}$    with     increasing     $x$    is
clearly visible.  Correspondingly,  the  partial concentration  of $x_{s,i}$
are expected to increase  with increasing  $x$. 
For $x>1.5\%$ we obtain $x_{s} \approx 0.8 x$ and
$x_{i} \approx 0.2 x$, in  good agreement with the density-functional
results \cite{Masek:2004_a}.

The  linear relations between  $x_{s}$, $x_{i}$,  and $x$  reflect the
fact that  the difference of  the formation energies of  Mn$_{\rm Ga}$
and Mn$_{\rm I}$ impurities (see right inset of Fig.~\ref{partial}) can be, up to $x = 10\%$, approximated by a
linear function of $x_{s}$ and $x_{i}$,
\begin{equation}
\Delta(x_{s},x_{i}) \equiv F_{s}(x_{s},x_{i}) - F_{i}(x_{s},x_{i}) 
\approx -0.1 + 5.9 x_{s} - 15.1 x_{i} ({\rm
eV}).
\end{equation}
This  relation allows us to interpret 
the theoretical 
distribution of Mn atoms between substitutional and
interstitial sites. For $x < 1.5\%$,
Mn$_{\rm  Ga}$  has a  lower  formation  energy than Mn$_{\rm I}$ 
and Mn  atoms  tend to occupy
 substitutional   positions.  At $x\approx 1.5\%$,
$\Delta(x_{s},x_{i})$ approaches zero and both Mn$_{\rm
Ga}$ and Mn$_{\rm I}$ are formed with a similar probability, 
as shown in Fig.~\ref{partial}.

We note that  both Mn$_{\rm  Ga}$ and  Mn$_{\rm I}$
positions remain metastable in  the whole concentration range shown in
Fig.~\ref{partial}  and that  our results  correspond to  the as-grown
rather  than  to  the  annealed  materials.  During  the  growth,  the
formation   energies   (namely   $\Delta(x_{s},x_{i})$)  control  
incorporation of Mn  atoms assuming that the total amount
of  Mn in  the material  is related  to a  sufficiently  high chemical
potential in the Mn source. The annealing processes, on the other hand, do not
depend  on formation  energies but rather on  energy barriers 
surrounding individual metastable positions of Mn in the lattice. 
The barriers are larger
for Mn$_{\rm Ga}$ \cite{Erwin:2002_a,Masek:2003_b} 
so that the post-grown low temperature annealing
can be used to remove Mn$_{\rm I}$ without changing the
number of Mn$_{\rm Ga}$ significantly.

\subsection{Hole density and Hall coefficient}
\label{theory_hall} 

As discussed above, the level of compensation is one of the key
parameters that determines Curie temperatures in (Ga,Mn)As.
In this paper, as well as in a number of other experimental
works, hole densities  are  obtained
from  Hall measurements.  In order  to
estimate  the  uncertainty of  this  experimental technique
we analyze in this section  theoretical Hall  factors,  $r_H=(\rho_{xy}-\rho_{xy,0})/(B/ep)$,
in ferromagnetic (Ga,Mn)As epilayers.  Here $\rho_{xy,0}$ is the Hall resistivity at
field $B=0$ which can be non-zero due to the anomalous Hall effect.

Detailed
microscopic  calculations  in   non-magnetic  p-type  GaAs  with  hole
densities $p\sim 10^{17}-10^{20}$~cm$^{-3}$ showed that $r_H$ can vary
between 0.87  and 1.75, depending on doping, scattering mechanisms, and
on the level on which the complexity of the GaAs valence band is
modelled.\cite{Kim:1995_a} Here we focus on estimating the effect on
$r_H$ of the  spin-splitting of the valence band  and of the anomalous
Hall term that is particularly large in ferromagnetic (Ga,Mn)As.

The calculations are based on numerical evaluation of the Kubo formula at   finite    magnetic   fields.   
We   assume band-   and    wavevector-independent quasiparticle  lifetimes for simplicity.  It is
essential for our analysis to allow for both intra-band and inter-band
transitions.  At  zero  magnetic  field,  the  inter-band  transitions
between SO-coupled,  spin-split bands give rise to  the anomalous Hall
effect (AHE), i.e., to a  non-zero $\rho_{xy}$ that is proportional to
the  magnetization.\cite{Sinova:2004_c} On the  other hand,  the ordinary  Hall resistance
which  is proportional to  $B$ arises,  within the  simple single-band
model, from intra-band transitions between adjacent Landau levels. The
Kubo formula that includes  both intra-band and inter-band transitions
allows us  to capture simultaneously 
the anomalous  and ordinary Hall effects in the
complex (Ga,Mn)As valence bands.

\begin{figure}
\hspace*{-.5cm}\includegraphics[angle=-0,width=3.8in]{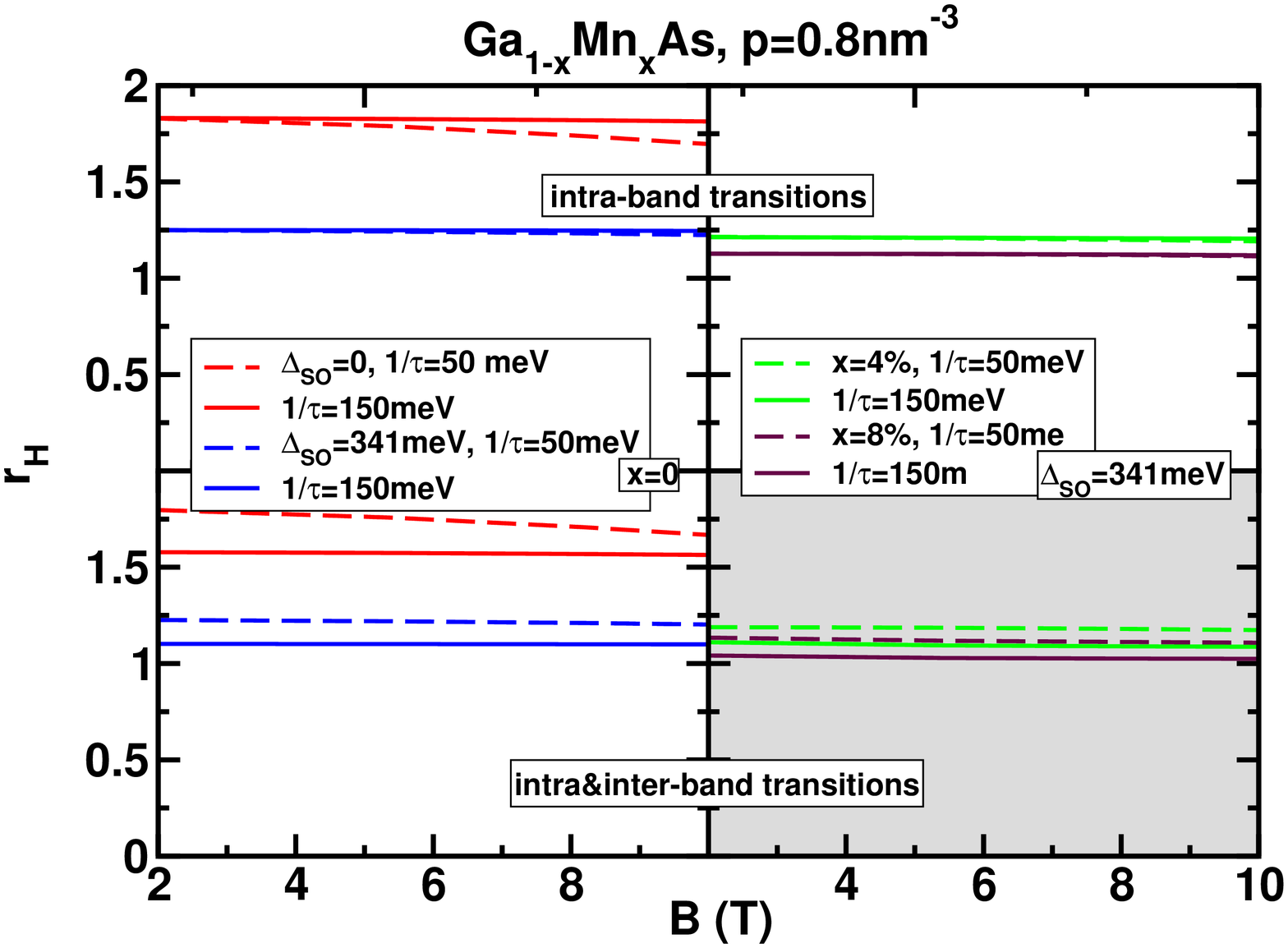}
\caption{Theoretical Hall factors for $p=0.8$~nm$^{-3}$; $\hbar/\tau=50$~meV (dashed lines), 
$\hbar/\tau=150$~meV (solid lines). Top panels: only intra-band transitions
are taken into account. Bottom panels: intra- and inter-band transitions are taken into account.
Left panels: GaAs ($x=0$); zero SO-coupling (red lines), $\Delta_{SO}=341$~meV (blue lines).
Right panels: (Ga,Mn)As with Mn$_{\rm Ga}$ concentration 4\% (green lines), 8\% (brown lines).
$\rho_{xy}=0$ in all panels except for the bottom left panel where $\rho_{xy}(B=0)\neq 0$ due
to the anomalous Hall effect.
} \label{rh8}
\end{figure}

\begin{figure}
\hspace*{-.5cm}\includegraphics[angle=-0,width=3.8in]{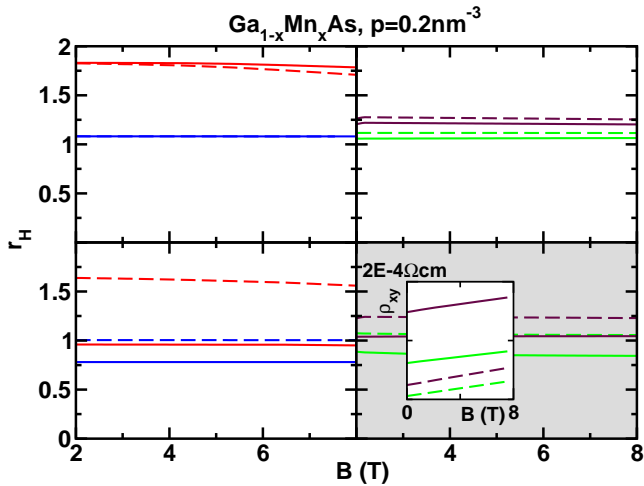}
\caption{Theoretical Hall factors for $p=0.2$~nm$^{-3}$; same line coding as in Fig.~\protect\ref{rh8}. 
Inset: Theoretical Hall curves showing the anomalous Hall effect contribution at $B=0$.
} \label{rh2}
\end{figure}

Many  of  the qualitative  aspects  of  the  numerical data  shown  in
Figs.~\ref{rh8} and \ref{rh2} can be explained using a simple model of
a conductor with two parabolic  uncoupled bands. Note that the typical
scattering rate in (Ga,Mn)As epilayers is $\hbar/\tau\sim 100$~meV and
the cyclotron energy at  $B=5$~T is $\hbar\omega\sim 1$~meV, i.e., the
system is in  the strong scattering limit, $\omega\tau\ll  1$. In this
limit, the two band model gives resistivities:
\begin{eqnarray}
\rho_{xx}&\approx&\frac{1}{\sigma_{xx,1}+\sigma_{xx,2}}
\approx\frac{1}{\sigma_{0,1}+\sigma_{0,2}}\nonumber \\
\rho_{xy}&\approx& -\frac{\sigma_{xy,1}+\sigma_{xy,2}}
{(\sigma_{xx,1}+\sigma_{xx,2})^2}\nonumber\\
&=&\frac{B}{ep_1}\;\frac{1+\frac{p_2}{p_1}(\frac{m^{\ast}_1}{m^{\ast}_2})^2}
{(1+\frac{p_2}{p_1}\frac{m^{\ast}_1}{m^{\ast}_2})^2}\ge\frac{B}{ep}\;, \label{rho_2band}
\label{hall_two_band}
\end{eqnarray}
where  the  indices  1 and  2  correspond  to  the  1st and  2nd  band
respectively,  the  total  density  $p=p_1+p_2$,  and  the  zero-field
conductivity   $\sigma_0=e^2\tau   p/m^{\ast}$.   Eq.~(\ref{rho_2band})
suggests that in the strong  scattering limit the multi-band nature of
the  hole  states   in  (Ga,Mn)As  should  not  result   in  a  strong
longitudinal  magnetoresistance. This  observation is  consistent with
the measured weak dependence of $\rho_{xx}$ on $B$ for magnetic fields
at which magnetization in the (Ga,Mn)As ferromagnet is saturated.\cite{Edmonds:2002_a}

The simple two-band  model also suggests that the  Hall factor, $r_H$,
is larger than one in multi-band systems with different dispersions of
individual  bands.  Indeed, for  uncoupled  valence  bands, i.e.  when
accounting  for  intra-band   transitions  only,  the  numerical  Hall
factors in  the top panels of Figs.~\ref{rh8}  and \ref{rh2} are
larger  than 1 and independent of $\tau$ as also suggested by 
Eq.~(\ref{hall_two_band}).  The  suppression of  $r_H$  when SO-coupling  is
turned on, shown in the  same graphs, results partly from depopulation
of  the   angular  momentum  $j=1/2$  split-off   bands.  In addition to  this
``two-band  model'' like  effect, the  inter-Landau-level matrix
elements are reduced  due to SO-coupling since the  spinor part of the
eigenfunctions   now  varies   with   the  Landau   level  index.   In
ferromagnetic Ga$_{1-x}$Mn$_x$As  the bands are  spin-split and higher
bands depopulated  as $x$  increases. In terms  of $r_H$,  this effect
competes with  the increase of the  inter-Landau-level matrix elements
since the  spinors are now more  closely aligned within a  band due to
the exchange  field produced by  the polarized Mn  moments. Increasing
$x$ can therefore lead to both decrease or increase of $r_H$ depending
on  other  parameters of  the  (Ga,Mn)As,  such  as the  hole  density
(compare top right panels of Figs.~\ref{rh8} and \ref{rh2}).

The inter-band transitions result in a more single-band like character
of  the  system,  i.e.  $r_H$   is  reduced,  and  the  slope  of  the
$\rho_{xy}(B)$ curve now depends more strongly on $\tau$. Although the
anomalous and ordinary Hall effect contributions to $\rho_{xy}$ cannot
be  simply decoupled,  the comparison  of numerical  data in  the four
panels and  the inset in Fig.~\ref{rh2} confirms  the usual assumption
that the  anomalous Hall  effect produces a  field-independent off-set
proportional to  magnetization and  $\rho_{xx}^2$. The comparison  also suggests
that  after  subtracting  $\rho_{xy}(B=0)$,   $r_H$  can  be  used  to
determine the hole  density in (Ga,Mn)As with accuracy  that is better
than in non-magnetic GaAs  with comparable hole densities. For typical
hole and Mn densities  in experimental (Ga,Mn)As epilayers we estimate
the error of the Hall measurement of $p$ to be within $\pm 20\%$.

\section{Experiment}

\label{exp}

\subsection{Measured Curie temperatures and hole densities}

\label{exp_tc} 

A series of (Ga,Mn)As thin films with varying Mn content were grown by
LT-MBE epitaxy using As$_{\rm  2}$. The layer
structure is 25 or 50nm (Ga,Mn)As / 50nm low temperature GaAs / 100nm
high temperature ($580^{\circ}$C) GaAs / GaAs(100) substrate. For a
given Mn  content, the growth  temperature of the (Ga,Mn)As  layer and
the GaAs buffer  is chosen in order to  minimize As antisite densities
while   maintaining  two-dimensional   growth  and   preventing  phase
segregation. We find that the  growth temperature must be decreased as
the  Mn concentration  is increased:  for  the lowest  Mn content  the
growth  temperature was $\sim  300^{\circ}$C, for  the highest  it was
$\sim  180^{\circ}$C.  Full  details   of  the  growth  are  presented
elsewhere.\cite{Campion:2003_a,Foxon:2004_a}

The  Mn content  was controlled  by  varying the  Mn/Ga incident  flux
ratio,  measured $in-situ$  and  calibrated using  SIMS measurements  on 1$\mu$m  thick  (Ga,Mn)As films,
grown under  otherwise identical conditions to  the samples considered
here. A  detailed comparison of the  results of a  number of different
calibration techniques, presented in Ref.~\onlinecite{Zhao:2005_b} allows us to
assign   an   uncertainty   of    $\pm$10$\%$   to   the   quoted   Mn
concentrations. However, it should be noted that the SIMS measurements
yield the $total$ volume density of Mn in the (Ga,Mn)As films, and not
the  fraction of  Ga substituted  by Mn.  This is  important as  it is
expected  that  a   fraction  of  the  Mn  will   be  incorporated  on
interstitial  as  well  as  substitutional  sites\cite{Yu:2002_a}.  We
define  the  Mn concentration,  $x$, as  the  total Mn
volume density relative to  the volume density of Ga in GaAs.

Hall  bar  structures, of  width  200 $\mu$m  and  length  1 mm,  were
fabricated       from      the      (Ga,Mn)As       samples      using
photolithography.  Simultaneous   magnetoresistance  and  Hall  effect
measurements   were   performed   using  standard   low-frequency   ac
techniques, in order  to extract both the Curie  temperature $T_c$ and
the  hole density $p$,  as detailed  below. Magnetic  fields of  up to
$\pm$0.7  T  and  $\pm$16.5 T  were  used  to  obtain $T_c$  and  $p$,
respectively. Following these  measurements, the samples were annealed
in  air at  $190^{\circ}$C.  The electrical  resistance was  monitored
during annealing, and the anneal was halted when this appeared to have
reached a minimum (after typically 50 to 150 hours). The $T_c$ and $p$
were then re-measured.

Below $T_c$, the Hall resistance $R_{xy}$ in (Ga,Mn)As is dominated by
the anomalous Hall  effect, with $R_{xy} \sim R_AM_z$,  where $M_z$ is
the perpendicular component of  the magnetization, and the coefficient
$R_A$  is  roughly proportional  to  the  square  of the  resistivity,
$\rho_{xx}$. Therefore,  $R_{xy}$ / $\rho_{xx}^2$ gives a  direct measurement of
$M_z$, which can  be used to extract $T_c$ using  Arrot plots.\cite{Matsukura:1998_a} The
value of $T_c$ obtained depends  only weakly on the precise dependence
of $R_A$ on  $\rho_{xx}$ assumed, since $\rho_{xx}$ varies  only slowly close to
$T_c$, while $R_{xy}$ varies rapidly.  We are therefore able to obtain
$T_c$ within an accuracy of $\pm$1~K by this method.\cite{Edmonds:2002_a}

\begin{figure}
\includegraphics[angle=-90,width=4.0in]{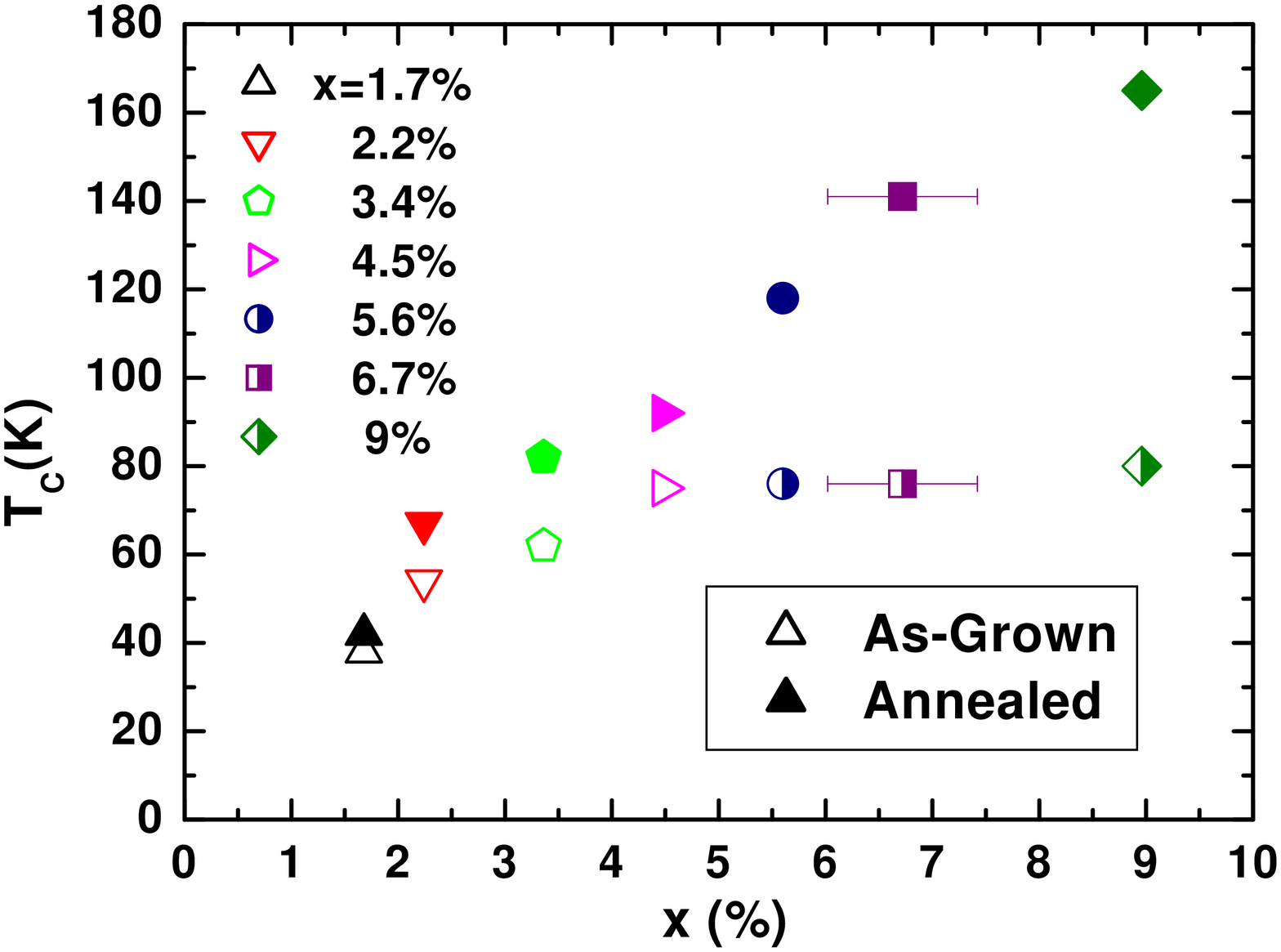}
\caption{
Experimental Curie temperature versus total Mn doping. $T_c$ is measured from the anomalous Hall effect, 
Mn doping by SIMS. Open symbols correspond to as-grown samples, half-open symbols to as-grown samples with large charge compensation, and filled symbols to annealed samples. For clarity, error bars are shown only for the x=6.7\% sample.
} 
\label{tc_Mn_exp}
\end{figure}

$T_c$ obtained  for the  (Ga,Mn)As Hall bar  samples before  and after
annealing are  shown versus  $x$ in Fig.~\ref{tc_Mn_exp}. It can  be seen
that  the  low-temperature annealing  procedure  results  in a  marked
increase in $T_c$ as has been found previously.\cite{Hayashi:2001_a} Increases of $T_c$
by more than a factor of  two are possible. This effect becomes larger
as  the Mn  concentration increases.  Since the  $T_c$ enhancement is
associated with out-diffusion and passivation of interstitial Mn,\cite{Edmonds:2004_a} this
indicates that  as the  incident Mn flux  is increased,  an increasing
fraction  is  incorporated  on  interstitial sites,  as  predicted  in
Section~\ref{theory_partial}.

\begin{figure}

\includegraphics[angle=-90,width=4.0in]{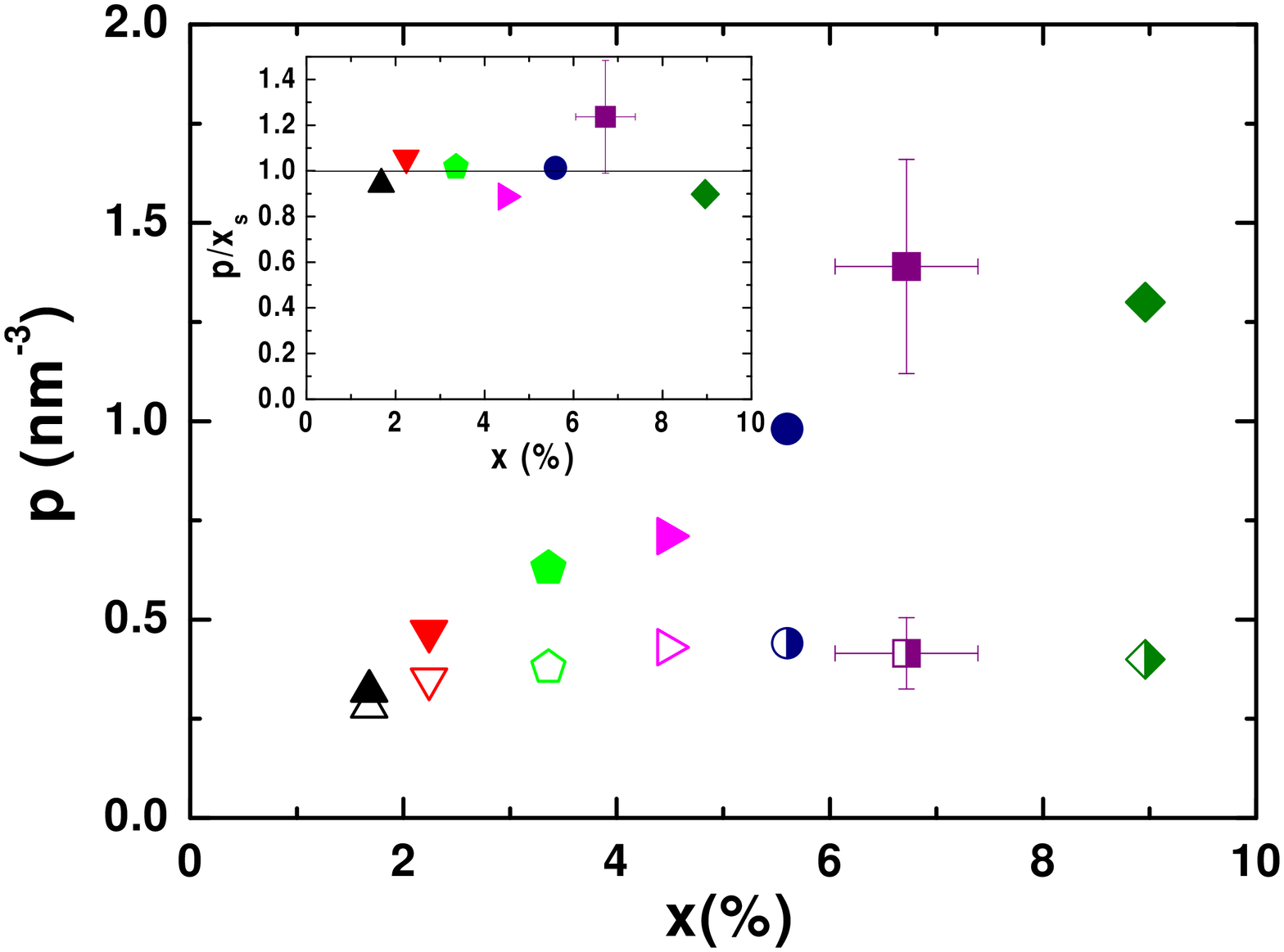}
\caption{
Experimental hole density versus total Mn doping. 
Hole density is measured by ordinary Hall effect. Same symbol coding is 
used as in Fig.\protect\ref{tc_Mn_exp}. 
}
\label{hole_Mn_exp}
\end{figure}

To obtain  hole densities from  $R_{xy}$, it is necessary  to separate
the  small  normal Hall  term  from  the  much larger  anomalous  Hall
term.  Measurements were  performed at  0.3 K  and in  magnetic fields
above 10 T, i.e., under conditions where the normal Hall term gives the
dominant field-dependent contribution  to $R_{xy}$. Then, the measured
$R_{xy}$ was fitted  to ($\alpha\rho^2 + r_HB$), where  $\rho_{xx}$ and $B$
are  the measured  resistivity and  magnetic field,  and  $\alpha$ and
$r_H$ are fit  parameters. Finally, the hole density  is obtained from
$r_H=1/(pew)$, where  $w$ is the  (Ga,Mn)As layer thickness.  From the
detailed  calculations described  in Section~\ref{theory_hall} we  can  ascribe an
uncertainty of $\pm$20\% to the values of  $p$ obtained using this
method. The measured $p$ for the (Ga,Mn)As Hall bar samples before and
after annealing are shown versus  $x$ in Fig.~\ref{hole_Mn_exp}. We see that
annealing greatly increases $p$ for large $x$.
Data in the inset of Fig.~\ref{hole_Mn_exp}, discussed in detail below, show that within error the 
samples are uncompensated after annealing.

\subsection{Substitutional and interstitial Mn}

\label{exp_partial} 

\begin{figure}
\includegraphics[angle=-90,width=4.0in]{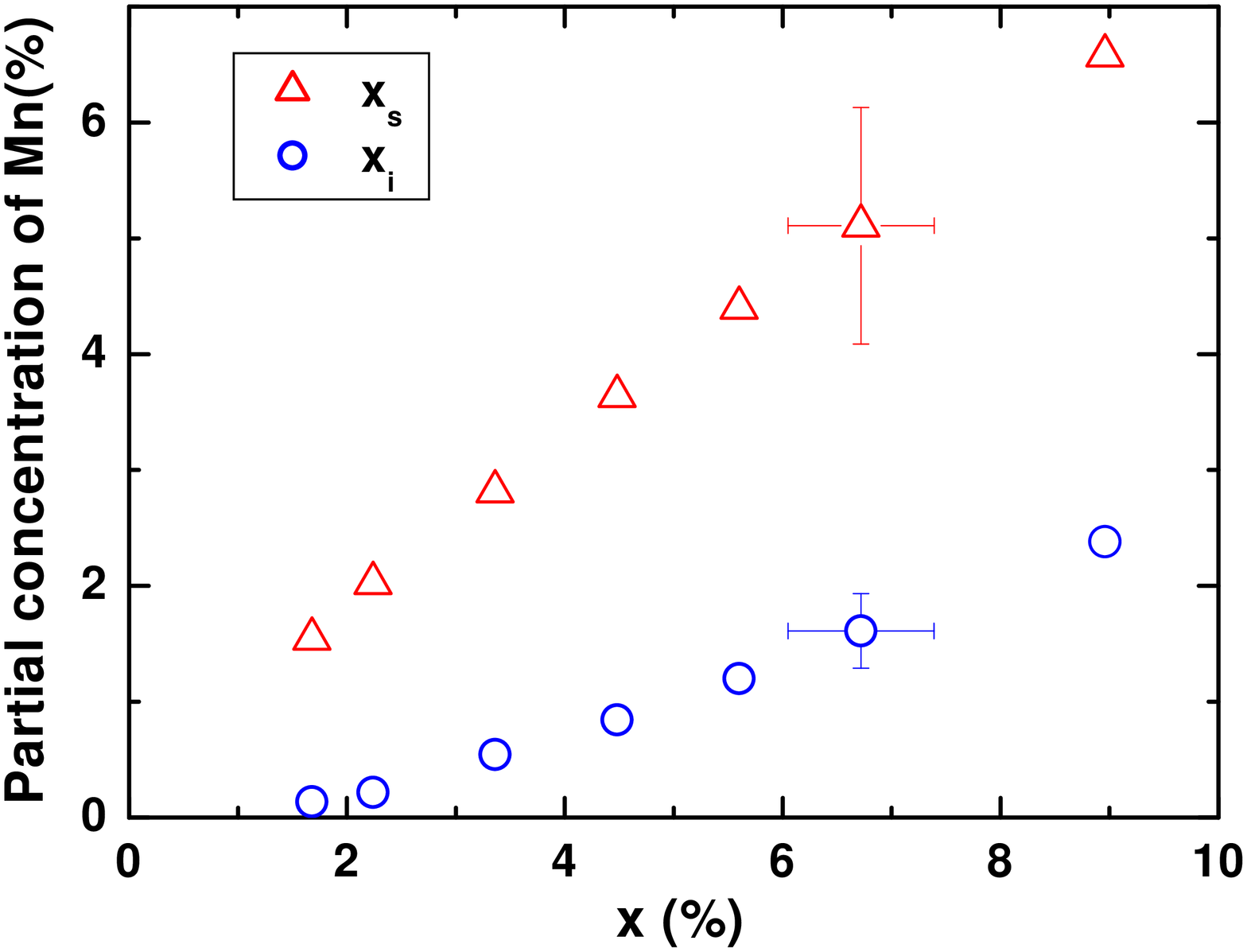}
\caption{
Experimental partial concentrations of Mn$_{\rm Ga}$ (triangles) and Mn$_{\rm I}$ (circles)
in as
grown samples. 
Data show no saturation of Mn$_{\rm Ga}$ with 
increasing total Mn doping consistent with theory expectation. 
}
\label{partial_exp}
\end{figure}

From the measured hole density $p$ before and after annealing, and the
total Mn density  $x$, values can be obtained  for the density
of incorporated Mn occupying  acceptor substitutional and double donor
interstitial lattice  sites, $x_{s}$  and $x_i$. These  are obtained
using the following assumptions: i) the only contribution to the total
Mn density determined by SIMS are from substitutional and interstitial
Mn,  i.e.  $x  =  x_{s}  +  x_i$; ii)  the  only  source  of
compensation in the (Ga,Mn)As films are the interstitial Mn, which are
double  donors i.e.  $p =  4/a_{lc}^3(x_{s} - 2 x_i)$; iii)  the  low temperature
annealing procedure affects only  $x_i$, and not $x_{s}$. The values
of $x_s$  and $x_i$ in  the unannealed films obtained  under these
assumptions are  shown in Fig.~\ref{partial_exp}. We  find a remarkably good agreement between
experiment and the theory data in Fig.~\ref{partial} and in Ref.\onlinecite{Masek:2004_a}. 
As a consistency check, we show in the inset of
Fig.~\ref{hole_Mn_exp} the ratio of  hole density to substitutional Mn$_{\rm Ga}$ density after
annealing,  as  obtained  under  the  above  assumptions.  Within  the
experimental  error we  obtain one  hole per  substitutional  Mn$_{\rm Ga}$ after
annealing,  that  is, there  is  no  significant  compensation in  the
annealed  (Ga,Mn)As films.  This justifies  our neglect  of additional
compensating defects  such as  As$_{\rm Ga}$ in  determining $x_s$
and $x_i$.

\subsection{${\bf T_c}$ versus Mn$_{\rm\bf Ga}$, effective Mn$_{\rm\bf Ga}$, and hole densities}

\label{exp_theory}

Since we obtain reasonably accurate values for $T_c$, hole densities and 
the partial Mn$_{\rm Ga}$  and Mn$_{\rm I}$ concentrations for the set of samples considered 
here, we now attempt to assess on the basis of the experimental data 
the key factors determining $T_c$ and to compare the experimental results 
with the broad predictions of theory. 

\begin{figure}
\includegraphics[angle=-90,width=4.0in]{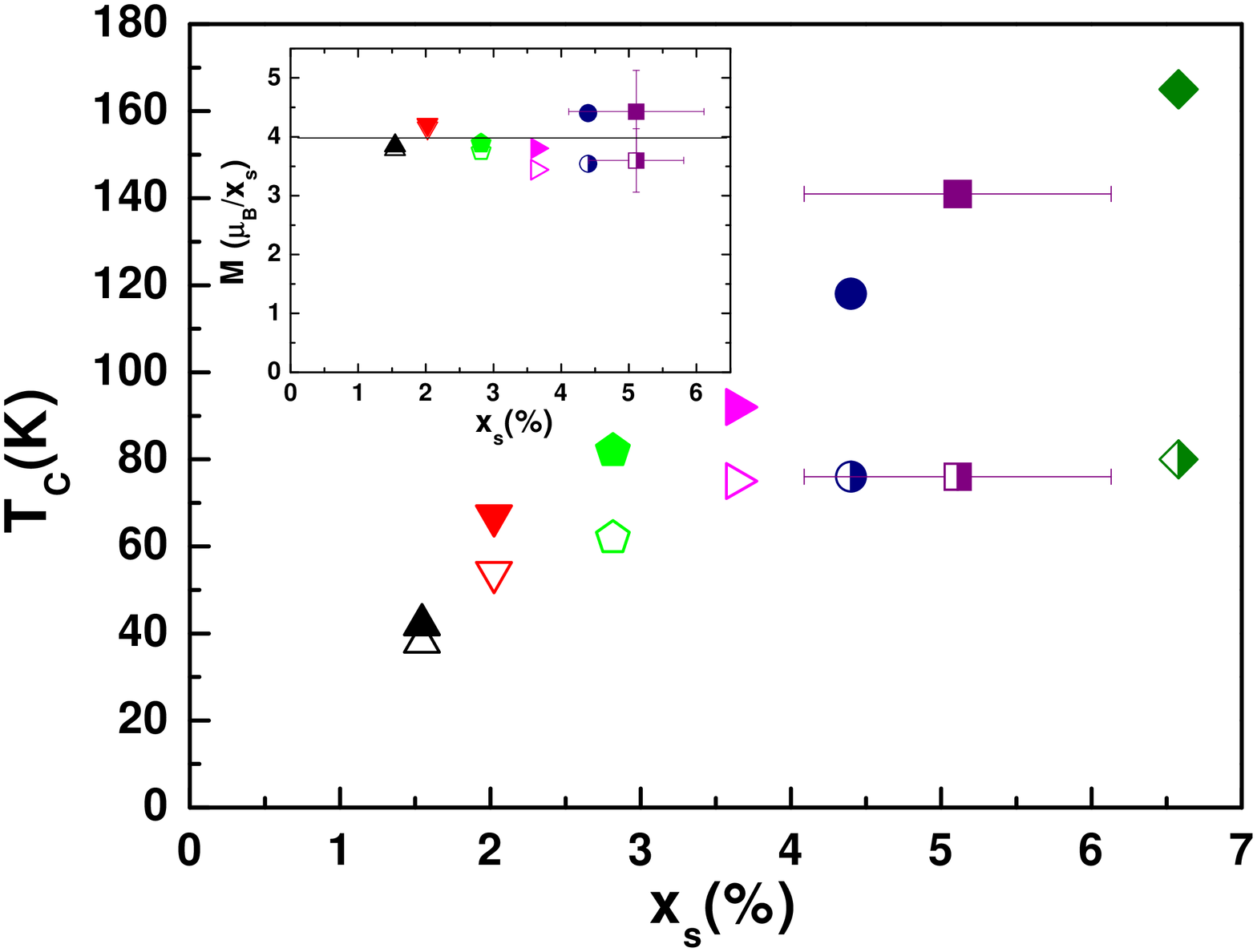}
\caption{  Experimental $T_c$  versus Mn$_{\rm  Ga}$ concentration, $x_s$ (see text for definition of $x_s$). Magnetization
per $x_{s}$ is shown in the inset.
}
\label{tc_Mn_sub_exp}
\end{figure}

\begin{figure}
\includegraphics[angle=-90,width=4.0in]{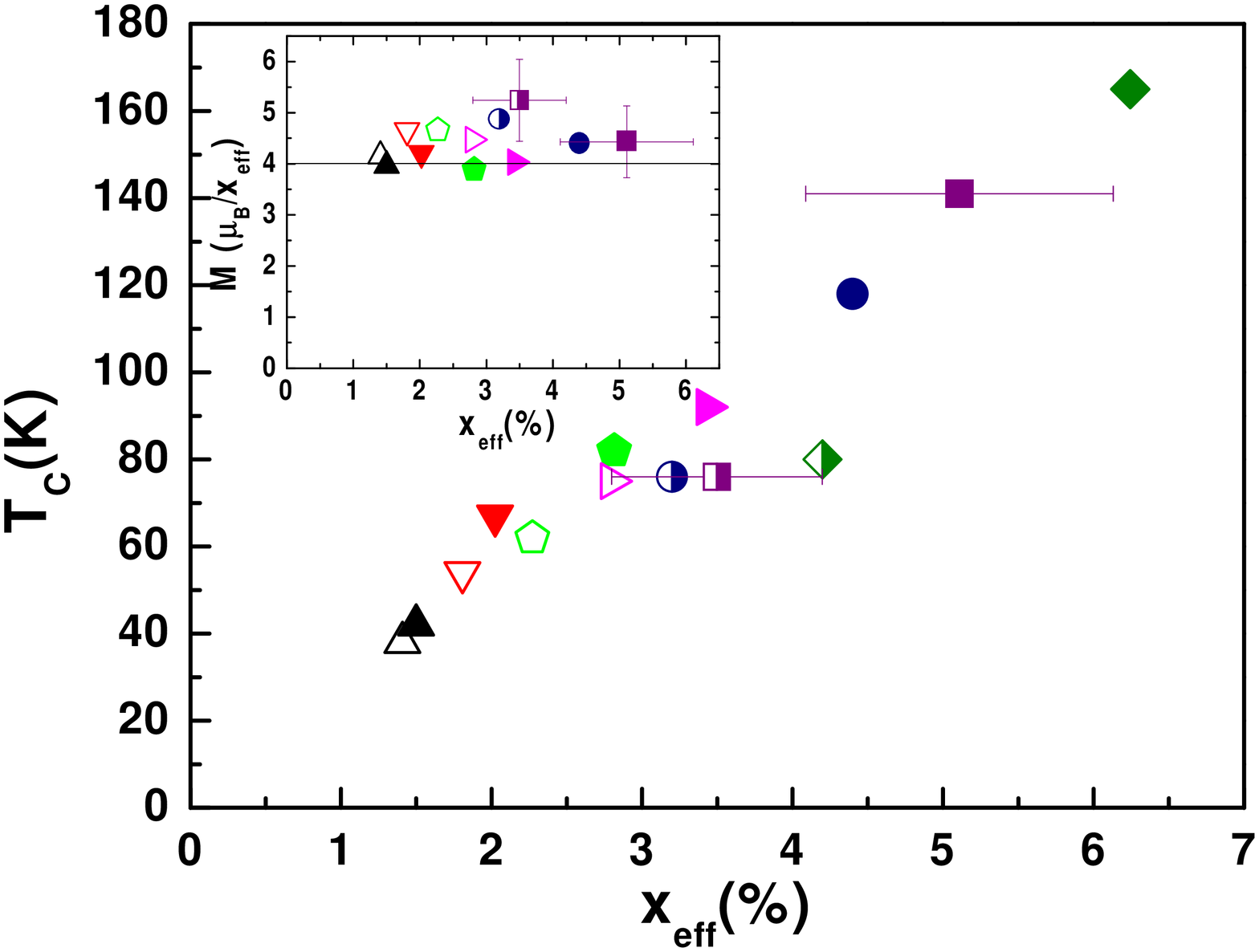}
\caption{ Experimental $T_c$ versus  effective Mn$_{\rm Ga}$ concentration, $x_{eff}$ 
(see text for definition of $x_{eff}$). Magnetization
per $x_{eff}$ is shown in the inset.
}
\label{tc_Mn_eff_exp}
\end{figure}

In Fig.~\ref{tc_Mn_exp} $T_c$ was plotted against total Mn concentration. Before 
annealing the $T_c$ values of samples with high compensation (samples with large 
compensation are indicated as half filled symbols in 
this and subsequent figures) do not increase significantly with increasing 
total Mn density but a steady increase is recovered after annealing. In Fig.~\ref{tc_Mn_sub_exp} 
$T_c$ is plotted against the substitutional Mn$_{\rm Ga}$ concentration. The form of 
Figs.~\ref{tc_Mn_exp} and  \ref{tc_Mn_sub_exp} are broadly similar despite the different x-axes. We expect, however, and
will assume in the following discussion that  any Mn$_{\rm I}$ donor present is attracted
to a Mn$_{\rm Ga}$ acceptor  and that the pair couples antiferromagnetically.\cite{Masek:2003_b} 
Then the effective uncompensated moment
density will be $x_{eff} = x_s -x_i$. 
Plotting $T_c$ against $x_{eff}$ in Fig.~\ref{tc_Mn_eff_exp} reveals that for all the low 
compensation samples $T_c$ increases approximately linearly with $x_{eff}$ but 
that as compensation, $(1-pa_{lc}^3/4x_{eff})$, increases above $\sim 40\%$ the measured $T_c$ 
values fall increasingly far below this linear trend. 

\begin{figure}
\includegraphics[angle=-90,width=4.0in]{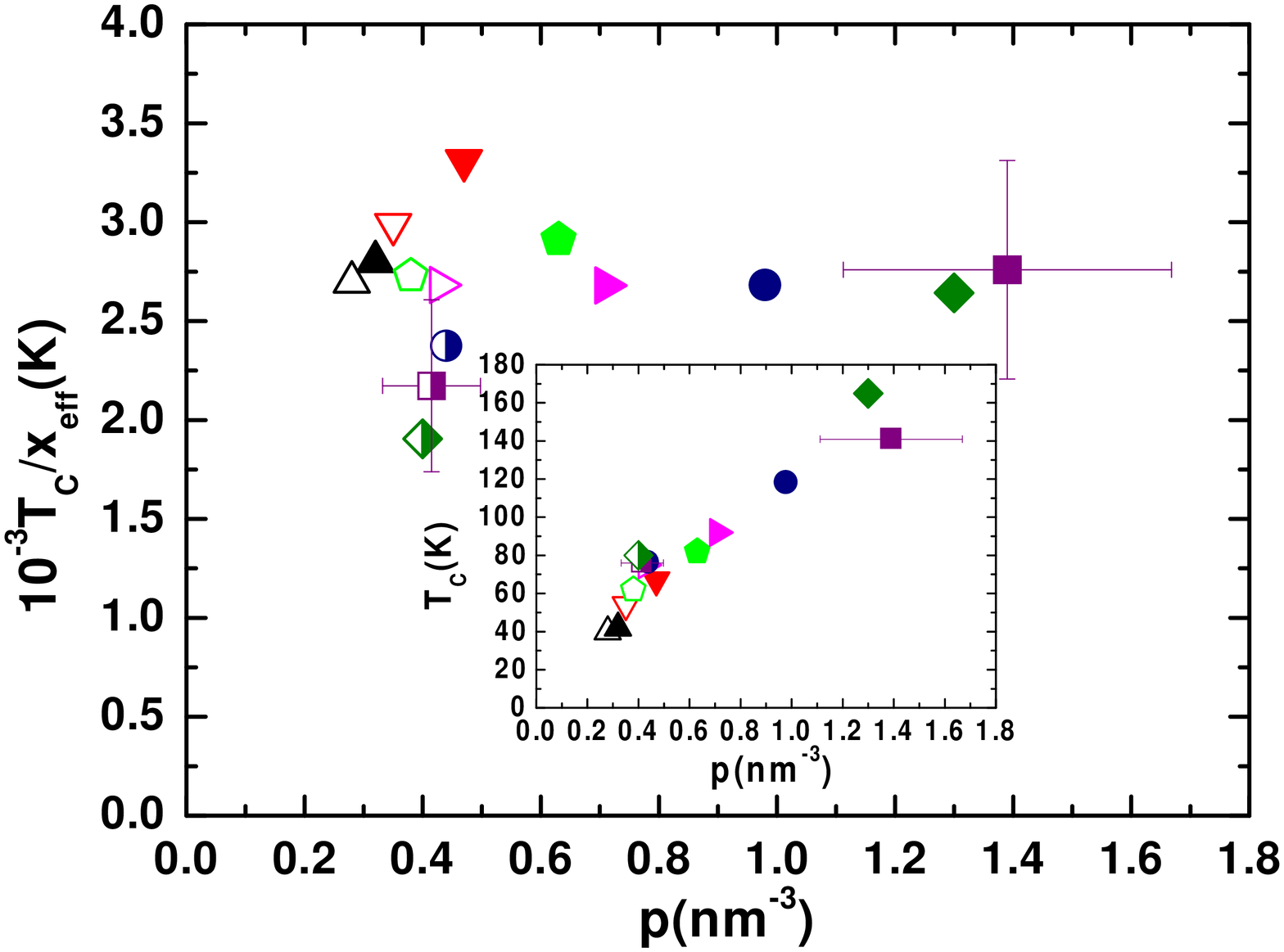}
\caption{
Experimental $T_c/x_{eff}$ versus hole density. $T_c/x_{eff}$ is nearly independent of hole density
except in highly compensated samples. Inset: $T_c$ versus hole density.
}
\label{tc_p_exp}
\end{figure}

\begin{figure}
\includegraphics[angle=-90,width=4.0in]{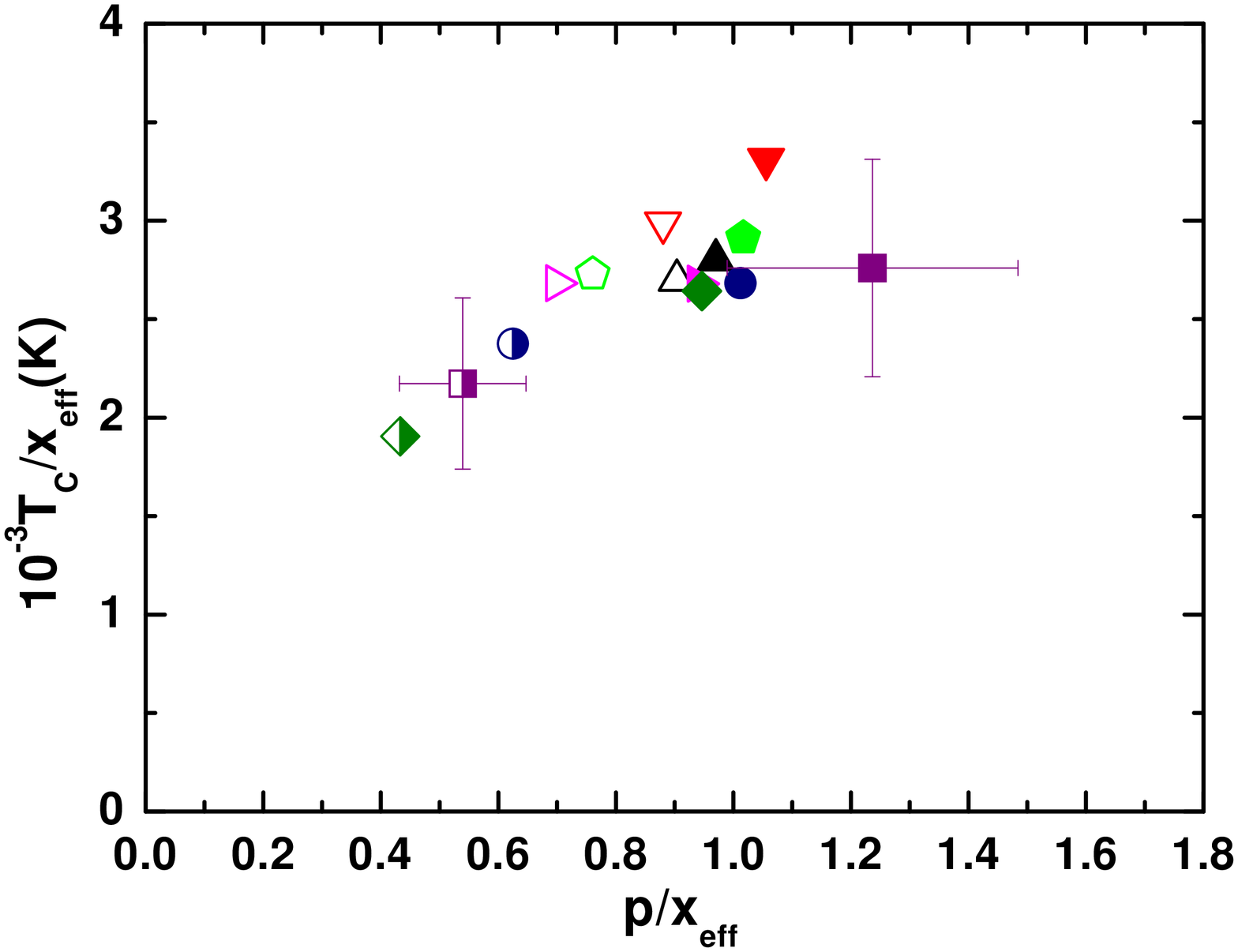}
\caption{ Experimental $T_c/x_{eff}$ versus hole density  relative to  effective concentration
of Mn$_{\rm Ga}$ moments. Deviation from linear dependence on $x_{eff}$ are seen only for
high compensations  ($1-pa_{lc}^3/4x_{eff}>40\%$) in agreement with  theory. For weakly
compensated samples $T_c$ shows  no signs of saturation with increasing
$x_{eff}$.  
}
\label{tc_p_Mn_eff_exp}
\end{figure}

If $T_c$ is plotted against  hole density, as is done in the inset of Fig.~\ref{tc_p_exp},
it is found to increase monotonically. However, this is primarily due to 
the increase in hole density with $x_{eff}$. The main plot in Fig.~\ref{tc_p_exp} shows that $T_c/x_{eff}$ is 
almost independent of hole density except for the case of the high 
compensation samples which clearly stand out as showing different behavior. 
To compare with the predictions of Section~\ref{theory} we finally plot $T_c/x_{eff}$ 
against $p/x_{eff}$ in Fig.~\ref{tc_p_Mn_eff_exp}. All experimental points in this plot show a common $T_c$ trend
and the magnitudes of 
the experimental and calculated $T_c/x_{eff}$ are comparable. Further confirmation of the theoretical picture 
is seen from the 
very weak experimental dependence of $T_c/x_{eff}$ on $p/x_{eff}$ for low compensation and 
the relatively rapid fall of $T_c/x_{eff}$ with decreasing  $p/x_{eff}$
for compensations of $\sim 40\%$ or larger.

As a consistency check for considering $x_{eff}$ as the density of local Mn$_{\rm Ga}$ 
moments participating in the ordered
ferromagnetic state, magnetization data are shown in insets of 
Figs.~\ref{tc_Mn_sub_exp} and \ref{tc_Mn_eff_exp}. 
Magnetizations were determined by superconducting quantum interference device magnetometry, 
at a sample temperature of 5K, and using an external field of 0.3~T to overcome in-plane anisotropy fields.
The 
charge and moment compensation after annealing is not significant for our samples and the moment 
per $x_s$  or $x_{eff}$ is within error 4-4.5~$\mu_B$. This  corresponds well to the 5~$\mu_B$ contribution
of  the S=5/2 local  
Mn$_{\rm Ga}$ moment and an approximately (-0.5)-(-0.8)~$\mu_B$ contribution of the antiferromagnetically 
coupled valence band hole\cite{Jungwirth:2005_a} in collinear (Ga,Mn)As ferromagnets. 
In the inset of Fig.~\ref{tc_Mn_sub_exp} we see that the measured moment per $x_s$ are 
all below 4~$\mu_B$ for the compensated samples. Including the effects of the 
Mn$_{\rm I}$-Mn$_{\rm Ga}$ antiferromagnetic coupling by considering the moment per $x_{eff}$ 
reveals again values around 4.5~$\mu_B$. Our conclusion therefore is that if we assume no significant frustration in 
our samples and account for the antiferromagnetic Mn$_{\rm I}$-Mn$_{\rm Ga}$ coupling, our extensive set of $T_c$, 
hole density, Mn density, and magnetization data brings up a clear common picture of $T_c$ trends in the  14
different (Ga,Mn)As ferromagnetic 
semiconductors we have studied, that is consistent with the theory predictions summarized in Section~\ref{theory}.

\section{Discussion}

\label{discussion} 
The preceding considerations of the  factors determining 
$T_c$ in (Ga,Mn)As lead us to conclude that there are no fundamental physics barriers 
to achieving room temperature ferromagnetism in this system. 
Experimental results for $T_c$  in samples in which  
compensating defects other than interstitial Mn have been reduced 
to very low levels have been shown to be in good agreement with 
theoretical expectations. Moment compensation by interstitial Mn$_{\rm I}$ impurities becomes 
increasingly important as the concentration of total Mn is increased. 
However, for the range of total Mn concentrations considered experimentally 
we find that the level of substitutional Mn$_{\rm Ga}$ continues to increase 
with $x$. Furthermore low temperature post growth 
annealing is found to effectively remove Mn$_{\rm I}$ in thin film 
samples even at large $x$, leading to material 
which within experimental error is both charge and moment uncompensated. Most importantly for 
samples in which the charge compensation is less than $\sim 40\%$  
we find theoretically and experimentally that $T_c$ increases approximately 
linearly with effective concentration, $x_{eff}$, of Mn$_{\rm Ga}$ whose moments are not compensated by near-neighbor Mn$_{\rm I}$ impurities. We have not observed any signs of saturation in this trend in the studied
(Ga,Mn)As diluted magnetic semiconductors. It should be noted that our 
maximum $x_{eff}$ is only 4.2\%  in the as grown 
sample and 6.2\%  after annealing for a total Mn concentration 
$x=9\%$. Hence the modest $T_c$'s observed so far. Achieving 
$T_c$ values close to room temperature in (Ga,Mn)As, which we expect to occur for  $x_{eff}\approx 10\%$ is essentially a 
technological issue, albeit a very challenging one. In the remaining paragraphs of this section we discuss these challenges in more detail. 

Low temperature MBE growth is used to achieve levels of Mn incorporation 
in (Ga,Mn)As far in excess of the equilibrium solubility level. When growing 
(Ga,Mn)As with Mn concentrations of several percent it is known that the 
Mn tends to accumulate on the surface\cite{Campion:2003_a,Foxon:2004_a}  in  a similar way  to all  high vapor
pressure  dopants  in GaAs  and  to  the higher  vapor  pressure
species, e.g. In in InGaAs. For homogenous Mn incorporation during
continuous  growth, a  surface Mn  concentration is  required  that is
higher than the bulk concentration.  For a given Mn concentration this
density gradient is  temperature dependent, increasing with increasing
temperature.  This leads to  an upper temperature limit for successful
growth  when the  Mn  surface concentration  approaches a  significant
proportion of a monolayer, after  which point surface clustering of Mn
occurs,  frustrating the  growth.\cite{Campion:2003_a,Foxon:2004_a}   Furthermore, higher  Mn fluxes
require lower growth temperatures.

The pursuit of higher Curie  temperatures has driven growth efforts to
very low temperatures compared with conventional MBE of GaAs.  In this
regime ($\sim$200 - $250^{\circ}$C) significant levels of compensating
defects such as As$_{\rm Ga}$ and vacancies usually occur in GaAs.\cite{Hurle:1999_a}  The density of As$_{\rm Ga}$  defects can be reduced by close to
stoichiometric growth  with As$_{\rm  2}$,\cite{Missous:1987_a} requiring  very precise
control over the As flux.

Apart from  precise control over  the stoichiometry and  the attendant
requirement for  flux stability,  a major technical  difficulty arises
from the measurement  and control of the growth  temperature. In order
to measure substrate temperature, most MBE machines in use today employ a
thermocouple  heated  by radiation  from  the  substrate or  substrate
holder. At normal growth temperatures ($\sim580^{\circ}$C) the radiant
flux from the substrate is high and the relationship between substrate
and thermocouple  is repeatable with  a short time  constant, allowing
for  good  temperature stability  and  control.  At low  temperatures,
however, the  radiant flux between  the substrate and  thermocouple is
low, leading  to a heightened  sensitivity to local conditions  such as
holder  emissivity,  radiant  heat  from the  metal  sources,  shutter
transients  etc., and  also  long time  constants. This  significantly
increases  the error  in the  temperature measurement  as well  as the
likelihood of temperature spikes and  drift as shutters are opened and
growth  proceeds.  In  MBE,   optical  pyrometers  are  ubiquitous  as
secondary  temperature  calibration   devices  but  most  cannot  read
accurately at these low temperatures and in many common configurations
suffer from  potential inaccuracies due  to reflection off the Knudsen
cells if used during growth.

It is  desirable to grow  at as high   temperature as possible  for a
given  Mn   flux,  while  maintaining   2D  growth  and   avoiding  Mn
clustering. However, with such  large errors and potential temperature
drift, growers  tend to err  towards lower than ideal  temperatures in
order to sustain the growth.   To explore fully  the parameter
space, effort should be directed  towards improving the control of both
metal fluxes and substrate temperature. 
This will maximize the chances
of  increasing the  doping towards  the  10\% 
Mn$_{\rm  Ga}$, required  for  room  temperature ferromagnetism. The 
increases in $T_c$ achieved in the last few years lead us to believe that 
higher transition temperatures will be obtained using conventional MBE. Growth interrupt strategies 
such as migration enhanced epitaxy (MEE)\cite{Sadowski:2001_a}  may have advantages 
over conventional MBE for the incorporation of higher levels of substitutional 
Mn however they will be especially sensitive to poor temperature stability 
and shutter transients and so will require even more precise temperature control.


\section{Conclusion}
\label{conclusion} 
Based on the broad agreement between theoretical and experimental Curie temperature trends in (Ga,Mn)As with 
Mn concentrations larger than 1.5\% we can outline the following strategies for achieving room 
temperature ferromagnetism in
this semiconductor:

(i) $T_c$ increases linearly with the concentration, $x_{eff}$, of local Mn$_{\rm Ga}$ 
moments participating in the ordered ferromagnetic state. Room temperature ferromagnetism 
should be achieved at $x_{eff}\approx 10\%$. Interstitial
Mn$_{\rm I}$ impurities reduce the number of these ordered Mn$_{\rm Ga}$ moments due to the strong antiferromagnetic
Mn$_{\rm Ga}$-Mn$_{\rm I}$ near-neighbor coupling. Mn$_{\rm I}$, however, can be 
efficiently removed by post-growth
annealing.
(ii) Equilibrium considerations, confirmed experimentally in samples with Mn$_{\rm Ga}$ concentrations up to 6.2\%,
suggest that  there is no fundamental physics barrier for increasing   Mn$_{\rm Ga}$ concentration to and beyond
10\%. A very precise control over the growth temperature and stoichiometry is, however,  
required for maintaing the 2D growth mode of the high quality, uniform (Ga,Mn)As materials.
(iii) Ferromagnetic coupling between the ordered local Mn$_{\rm Ga}$ moments is mediated by itinerant holes. For
charge compensations $(1-pa_{lc}^3/4x_{eff})>40\%$, the Curie 
temperature falls down with decreasing $p$. At compensations
smaller than $\sim 40\%$, however, $T_c$ is almost independent of the hole density. A modest  charge
compensation is, therefore, not an important limiting factor in the search  of high Curie temperature
ferromagnetic semiconductors based on (Ga,Mn)As and may 
be desirable to maximize the possibilities for doping and gate control of ferromagnetism.

\section*{Acknowledgment}
We acknowledge support from the Grant Agency of the Czech Republic through Grant No.  202/05/0575 and
Academy of Sciences of
the  Czech  Republic  through 
Institutional Support No. AV0Z10100521, from  the  EU  FENIKS  project
EC:G5RD-CT-2001-00535, the support from the UK EPSRC through Grant GR/S81407/01,
from the Welch Foundation,  the Department of Energy under grant DE-FG03-02ER45958,
and from Deutsche
Forschungsgemeinschaft through grant SFB 491.

\end{document}